\documentclass[aps,prl,preprint,tightenlines,superscriptaddress,showpacs,byrevtex]{revtex4}
\usepackage{graphicx}
\usepackage{dcolumn}
\usepackage{bm}
\usepackage{amsmath}
\usepackage{amssymb}
\usepackage{color}

\begin{document}

\preprint{\vbox{ 
								 \hbox{BELLE-CONF-0822}
                 \hbox{BELLE Preprint 2009-7}
                 \hbox{KEK Preprint 2008-56}
%                 \hbox{final v2.2 Apr. 2nd}
                              }}

\title{\quad\\
[0.5cm] Measurement of the Differential Branching Fraction 
and Forward-Backward Asymmetry for $B \to K^{(*)} \ell^+ \ell^-$}

\title{\quad\\
[0.5cm] Measurement of the Differential Branching Fraction 
and Forward-Backward Asymmetry for $B \to K^{(*)} \ell^+ \ell^-$}

%%% Paper:    B -> K(*) l+ l-
%%% Journal:  Physical Review Letters
%%% Contacts: J.-T. Wei (alan1215@hep1.phys.ntu.edu.tw)
%%% Non-responding authors or those who said NO are commented out.
%%% ====================================================================
%%% Click the RELOAD button on your web browser to see the updated file.
%%% ====================================================================
%%% Use \input{author} to insert this material into your latex file.
%%%%% Force institutions to appear in alphabetical order when typeset.
\affiliation{Budker Institute of Nuclear Physics, Novosibirsk}
%%%\affiliation{Chiba University, Chiba}
\affiliation{University of Cincinnati, Cincinnati, Ohio 45221}
\affiliation{T. Ko\'{s}ciuszko Cracow University of Technology, Krakow}
%%%\affiliation{Department of Physics, Fu Jen Catholic University, Taipei}
\affiliation{Justus-Liebig-Universit\"at Gie\ss{}en, Gie\ss{}en}
\affiliation{The Graduate University for Advanced Studies, Hayama}
%%%\affiliation{Gyeongsang National University, Chinju}
\affiliation{Hanyang University, Seoul}
\affiliation{University of Hawaii, Honolulu, Hawaii 96822}
\affiliation{High Energy Accelerator Research Organization (KEK), Tsukuba}
%%%\affiliation{Hiroshima Institute of Technology, Hiroshima}
%%%\affiliation{University of Illinois at Urbana-Champaign, Urbana, Illinois 61801}
\affiliation{Institute of High Energy Physics, Chinese Academy of Sciences, Beijing}
\affiliation{Institute of High Energy Physics, Vienna}
\affiliation{Institute of High Energy Physics, Protvino}
%%%\affiliation{Institute of Mathematical Sciences, Chennai}
%%%\affiliation{INFN - Sezione di Torino, Torino}
\affiliation{Institute for Theoretical and Experimental Physics, Moscow}
\affiliation{J. Stefan Institute, Ljubljana}
\affiliation{Kanagawa University, Yokohama}
\affiliation{Korea University, Seoul}
%%%\affiliation{Kyoto University, Kyoto}
\affiliation{Kyungpook National University, Taegu}
\affiliation{\'Ecole Polytechnique F\'ed\'erale de Lausanne (EPFL), Lausanne}
\affiliation{Faculty of Mathematics and Physics, University of Ljubljana, Ljubljana}
\affiliation{University of Maribor, Maribor}
\affiliation{University of Melbourne, School of Physics, Victoria 3010}
\affiliation{Nagoya University, Nagoya}
\affiliation{Nara Women's University, Nara}
\affiliation{National Central University, Chung-li}
\affiliation{National United University, Miao Li}
\affiliation{Department of Physics, National Taiwan University, Taipei}
\affiliation{H. Niewodniczanski Institute of Nuclear Physics, Krakow}
\affiliation{Nippon Dental University, Niigata}
\affiliation{Niigata University, Niigata}
\affiliation{University of Nova Gorica, Nova Gorica}
\affiliation{Novosibirsk State University, Novosibirsk}
\affiliation{Osaka City University, Osaka}
%%%\affiliation{Osaka University, Osaka}
%%%\affiliation{Panjab University, Chandigarh}
%%%\affiliation{Peking University, Beijing}
%%%\affiliation{Princeton University, Princeton, New Jersey 08544}
\affiliation{RIKEN BNL Research Center, Upton, New York 11973}
\affiliation{Saga University, Saga}
\affiliation{University of Science and Technology of China, Hefei}
\affiliation{Seoul National University, Seoul}
%%%\affiliation{Shinshu University, Nagano}
\affiliation{Sungkyunkwan University, Suwon}
\affiliation{University of Sydney, Sydney, New South Wales}
%%%\affiliation{Tata Institute of Fundamental Research, Mumbai}
\affiliation{Toho University, Funabashi}
\affiliation{Tohoku Gakuin University, Tagajo}
\affiliation{Tohoku University, Sendai}
\affiliation{Department of Physics, University of Tokyo, Tokyo}
%%%\affiliation{Tokyo Institute of Technology, Tokyo}
\affiliation{Tokyo Metropolitan University, Tokyo}
\affiliation{Tokyo University of Agriculture and Technology, Tokyo}
%%%\affiliation{Toyama National College of Maritime Technology, Toyama}
\affiliation{IPNAS, Virginia Polytechnic Institute and State University, Blacksburg, Virginia 24061}
\affiliation{Yonsei University, Seoul}
  \author{J.-T.~Wei}\affiliation{Department of Physics, National Taiwan University, Taipei} % Taiwan
  \author{P.~Chang}\affiliation{Department of Physics, National Taiwan University, Taipei} % Taiwan
  \author{I.~Adachi}\affiliation{High Energy Accelerator Research Organization (KEK), Tsukuba} % KEK
  \author{H.~Aihara}\affiliation{Department of Physics, University of Tokyo, Tokyo} % Tokyo
% \author{K.~Arinstein}\affiliation{Budker Institute of Nuclear Physics, Novosibirsk}\affiliation{Novosibirsk State University, Novosibirsk} % BINP
% \author{T.~Aso}\affiliation{Toyama National College of Maritime Technology, Toyama} % Toyama
  \author{V.~Aulchenko}\affiliation{Budker Institute of Nuclear Physics, Novosibirsk}\affiliation{Novosibirsk State University, Novosibirsk} % BINP
  \author{T.~Aushev}\affiliation{\'Ecole Polytechnique F\'ed\'erale de Lausanne (EPFL), Lausanne}\affiliation{Institute for Theoretical and Experimental Physics, Moscow} % ITEP
% \author{T.~Aziz}\affiliation{Tata Institute of Fundamental Research, Mumbai} % Tata
% \author{S.~Bahinipati}\affiliation{University of Cincinnati, Cincinnati, Ohio 45221} % Cincinnati
  \author{A.~M.~Bakich}\affiliation{University of Sydney, Sydney, New South Wales} % Sydney
  \author{V.~Balagura}\affiliation{Institute for Theoretical and Experimental Physics, Moscow} % ITEP
% \author{Y.~Ban}\affiliation{Peking University, Beijing} % Peking
  \author{E.~Barberio}\affiliation{University of Melbourne, School of Physics, Victoria 3010} % Melbourne
% \author{A.~Bay}\affiliation{\'Ecole Polytechnique F\'ed\'erale de Lausanne (EPFL), Lausanne} % Lausanne
% \author{I.~Bedny}\affiliation{Budker Institute of Nuclear Physics, Novosibirsk}\affiliation{Novosibirsk State University, Novosibirsk} % BINP
% \author{K.~Belous}\affiliation{Institute of High Energy Physics, Protvino} % Protvino
% \author{V.~Bhardwaj}\affiliation{Panjab University, Chandigarh} % Panjab
% \author{S.~Blyth}\affiliation{National United University, Miao Li} % NUU
  \author{A.~Bondar}\affiliation{Budker Institute of Nuclear Physics, Novosibirsk}\affiliation{Novosibirsk State University, Novosibirsk} % BINP
% ...
% BINP
\collaboration{The Belle Collaboration}

\begin{abstract}
We study $B \to K^{(*)} \ell^+ \ell^-$ decays ($\ell = e, \mu$) based on a data sample 
of 657 million $B\overline{B}$ pairs collected with the Belle detector at the KEKB $e^+e^-$ collider. 
We report the differential branching fraction, isospin asymmetry, 
$K^{*}$ polarization, and the forward-backward asymmetry ($A_{FB}$) 
as functions of $q^2=M_{\ell\ell}^{2} c^2$.
%$q^2$ (dilepton invariant mass squared) are reported. 
%The fitted $A_{FB}$ spectrum tends to be higher than 
%the Standard Model expectation by 2.7$\sigma$ difference.
The fitted $A_{FB}$ spectrum 
exceeds the Standard Model expectation by 2.7 standard deviations.
The measured branching fractions are 
%and lepton flavor ratios (muon/electron) are 
$\mathcal{B}$($B \to K^* \ell^+ \ell^-$) = (10.7$^{+1.1}_{-1.0}\pm0.9) \times 10^{-7}$
and 
$\mathcal{B}$($B \to K \ell^+ \ell^-$) = (4.8$^{+0.5}_{-0.4}\pm0.3) \times 10^{-7}$, 
where the first errors are statistical and the second are systematic, 
with the muon to electron ratios 
$R_{K^*}$ = 0.83$\pm0.17\pm0.08$ and $R_{K}$ = 1.03$\pm0.19\pm0.06$, respectively. 

\end{abstract}

%\collaboration{Belle Collaboration} \noaffiliation

\maketitle
    
\tighten
The $b \to s \ell^+\ell^-$ transition is a flavor-changing neutral current (FCNC) process, 
which, in the Standard Model (SM), 
proceeds at lowest order via either a $Z$/$\gamma$ penguin diagram or a $W^+ W^-$ box diagram. 
%The effective Wilson coefficients $C_7$, $C_9$, and $C_{10}$ 
%describe the amplitudes from the electromagnetic penguin, 
%the vector electroweak, and the axial-vector electroweak contributions, respectively. 
%The transition can provides a stringent test on the SM 
Since their amplitudes may interfere with the contributions from non-SM particles~\cite{nonsm}, 
the transition can probe the presence of yet unobserved particles and processes. 
More specifically, the lepton forward-backward asymmetry ($A_{FB}$) 
and the differential branching fraction 
as functions of dilepton invariant mass ($M_{\ell\ell}$) 
in the decays $B \to K^{*} \ell^+\ell^-$ 
differ from the SM expectations in various extended models~\cite{kll_th}.
%are widely predicted in various models~\cite{kll_th}. 
The former is largely insensitive to 
the theoretical uncertainties of the form factors describing the decay, 
and can hence provide a stringent experimental test of the SM. 
The latter has been so far determined only with a modest precision~\cite{kll_exp_bl,kll_exp_bb}. 
It can be used to extract the information on 
the coefficients associated with the theoretical models as well.
%Therefore, 
%The differential branching fraction and the lepton forward-backward asymmetry ($A_{FB}$)
%as functions of dilepton invariant mass ($M_{\ell\ell}$) 
%in the decays $B \to K^{*} \ell^+\ell^-$ provide information on the 
%coefficients associated with various models~\cite{kll_th}. 
%and can probe for the presence of new physics.

In this paper, we report measurements of 
the differential branching fractions and the isospin asymmetries 
as functions of $q^2=M_{\ell\ell}^{2} c^2$ for 
$B \to K^{*} \ell^+\ell^-$ and $B \to K \ell^+\ell^-$ decays.
The $K^{*}$ polarization and $A_{FB}$ for $B \to K^{*} \ell^+\ell^-$ decays 
as functions of $q^2$ are presented as well.
A data sample of 657 million $B\overline{B}$ pairs, 
corresponding to 605 fb$^{-1}$, 
collected with the Belle detector at the KEKB asymmetric-energy 
$e^+e^-$ collider~\cite{KEKB} is examined. 
The Belle detector is described in detail elsewhere~\cite{belle_detector}.
%Charge-conjugate decays are implied throughout the paper. 
%Charge-conjugate decays are implied and 
%equal production of $B^0\overline{B}{}^0$ and $B^+B^-$ pairs is assumed 
%throughout the paper.

We reconstruct $B \to K^{(*)} \ell^+\ell^-$ signal events in 10 final states: 
$K^{+} \pi^{-}$, $K_{S}^{0} \pi^{+}$, $K^{+} \pi^{0}$, $K^+$, 
and $K_{S}^{0}$ for $K^{(*)}$~\cite{chrg_conj}, 
combined with either electron or muon pairs.
All charged tracks other than the $K_{S}^{0}\to\pi^+\pi^-$ daughters 
are required to be associated with the interaction point (IP). 
%are required to have a maximum distance to the interaction point (IP) 
%of 5 cm along the beam direction ($z$) and 0.5 cm in the transverse plane ($r$--$\phi$).
%KID
%A track is identified as a $K^{+}$ ($\pi^{+}$) if 
%the kaon likelihood ratio is larger (less) than 0.6 (0.4);
%the kaon likelihood ratio is defined as 
%$\mathcal{R}_{K}\equiv \mathcal{L}_{K}/(\mathcal{L}_{K}+\mathcal{L}_{\pi})$,
%where $\mathcal{L}_{K}$ ($\mathcal{L}_{\pi }$) denotes a likelihood
%that combines measurements from the aerogel threshold Cherenkov counters, 
%the time-of-flight scintillation counters, 
%and $dE/dx$ from the central drift chamber 
%for the $K^{+}$ ($\pi^{+}$) tracks.
%This selection is about 85\% (89\%) efficient for kaons (pions) while 
%removing about 97\% (91\%) of pions (kaons). 
A track is identified as a $K^{+}$ ($\pi^{+}$) 
by combining information from the aerogel Cherenkov and time-of-flight subsystems 
with $dE/dx$ measurements in the central drift chamber~\cite{pid}.
%The kaon (pion) identification is \textcolor{blue}{90.1\%$\pm$1.0\% (93.5\%$\pm$0.6\%)} efficient 
%while removing \textcolor{blue}{92.7\%$\pm$0.4\% (93.3\%$\pm$1.7\%)} of pions (kaons).
The kaon (pion) identification is 
more than 85\% (89\%) efficient 
while removing more than 92\% (91\%) of pions (kaons).
%LID
For muon (electron) candidates, 
we use the lepton identification likelihood described in Ref.~\cite{pid}, 
which we denote by $\mathcal{R}_{x}$ ($x$ denotes $\mu$ or $e$). 
%we define the likelihood ratio $\mathcal{R}_{x}$ ($x$ denotes $\mu$ or $e$) as
%$\mathcal{R}_{x}\equiv \mathcal{L}_{x}/({\mathcal{L}_{x}}+{\mathcal{L}_{{\rm not-}x}})$,
%where $\mathcal{L}_{x}$ and $\mathcal{L}_{{\rm not-}x}$ 
%are the likelihood measurements from the relevant detectors~\cite{lid}.
We select $\mu^\pm$ candidates with $p_{\mu}>$ 0.7~GeV/$c$ 
and momentum-dependent $\mathcal{R}_{\mu}$ requirements 
that retain (93.4\%$\pm$2.0\%) of muons while removing (98.8\%$\pm$0.2\%) of pions.
Electron candidates are required to have 
$\mathcal{R}_{e} >$ 0.9, $\mathcal{R}_{\mu} <$ 0.8, and $p_{e}>$ 0.4~GeV/$c$.
These requirements retain (92.3\%$\pm$1.7\%) of electrons while removing (99.7\%$\pm$0.1\%) of pions.
Bremsstrahlung photons emitted by electrons are recovered by adding 
neutral clusters found within a 50 mrad cone along the electron direction. 
The cluster energies are required to be 
between 20 and 500~MeV.
%less than 0.5~GeV. 

Pairs of oppositely-charged tracks are used to reconstruct $K_{S}^{0} \to \pi^+ \pi^-$ candidates.
The invariant mass is required to lie within the range 483--513~MeV/$c^2$ 
($\pm$5 times the $K_{S}^{0}$ reconstructed-mass resolution). 
Other selection criteria are based on 
the distance and the direction of the $K_{S}^{0}$ vertex 
and the distance of daughter tracks to the IP.
For $\pi^{0} \to \gamma\gamma$ candidates, 
a minimum photon energy of 50~MeV is required 
and the invariant mass must be in the range 115~MeV/$c^2$ $< M_{\gamma\gamma} < $ 152~MeV/$c^2$ 
($\pm$3 times the $\pi^0$ reconstructed-mass resolution).
%%pi0 spectrum from MC = $134.6 \pm 11.42$~MeV
Requirements on the photon energy asymmetry 
$|E_{\gamma}^1 - E_{\gamma}^2|/(E_{\gamma}^1 + E_{\gamma}^2)<0.9$ 
%and the minimum momentum of the $\pi^0$ candidate in the lab frame, 
and $\pi^0$ momentum $p_{\pi^0} >$ 200~MeV/$c$ suppress the combinatorial background. 

$B$-meson candidates are reconstructed by combining a $K^{(*)}$ candidate and a pair of 
oppositely charged leptons, and selected using 
the beam-energy constrained mass $M_{\mathrm{bc}} \equiv \sqrt{E_{\mathrm{beam}}^{2} - p_{B}^{2}}$ 
and the energy difference $\Delta E \equiv E_{B} - E_{\mathrm{beam}}$, 
where $E_{B}$ and $p_{B}$ are the reconstructed energy and momentum
of the $B$ candidate in the $\Upsilon(4S)$ rest frame 
and $E_{\mathrm{beam}}$ is the beam energy in this frame. 
Bremsstrahlung photons are included in the calculation of the momenta of electrons 
and hence are included in the calculations of $M_{\mathrm{bc}}$, $E_{\mathrm{beam}}$ and $q^2$.
We require $B$-meson candidates to be within the region
$M_{\mathrm{bc}}>5.20$~GeV/$c^2$ and $-$35 ($-$55)~MeV $<\Delta E< 35$~MeV for the muon (electron) modes.
The signal region is defined as 5.27~GeV/$c^2$ $< M_{\mathrm{bc}} <$ 5.29~GeV/$c^2$. 
For the $K^*$ modes, the $M_{K\pi}$ candidate (signal) region is defined as 
$M_{K\pi} <$ 1.2~GeV/$c^2$ ($|M_{K\pi}-m_{K^*}|<80$~MeV/$c^2$).
%Here and in the rest of the paper 
%$M_X$ represents the invariant mass of the $X$ system and 
%$m_Y$ the nominal mass of $Y$.

The main backgrounds are continuum $e^+e^- \to q\overline{q}$ ($q=u,d,c,s$) 
and semileptonic $B$ decay events. 
We use the same set of variables and the likelihood ratio, $\mathcal{R}$, 
described in Ref.~\cite{pill}, for continuum $e^+e^- \to q\overline{q}$ suppression.
%A Fisher discriminant including 16 modified Fox-Wolfram moments~\cite{sfw} 
%is used to exploit the differences between
%the event shapes for continuum $q\overline{q}$ production 
%(jet-like) and for $B\overline{B}$ decay (spherical) in the $e^+e^-$ rest frame.
%We combine 1) a Fisher discriminant including 16 odified Fox-Wolfram moments~\cite{sfw}, 
%2) the missing mass $M_{\rm miss} \equiv \sqrt{E_{\rm miss}^{2} - p_{\rm miss}^{2}}$, 
%3) the angle between the reconstructed $B$ candidate and the beam direction ($\cos\theta_B$), 
%and 4) the distance in the $z$ direction between the candidate $B$ vertex and a vertex position 
%formed by the charged tracks that are not associated with the candidate $B$-meson 
%into a single likelihood ratio
%$\mathcal{R} = {\cal L}_s/({\cal L}_s + {\cal L}_{q\overline{q}})$, where
%${\cal L}_s$ (${\cal L}_{q\overline{q}}$) denotes the signal (continuum) likelihood.
For the suppression of semileptonic $B$ decays, 
%we combine the Fisher discriminant, missing mass, $\cos\theta_B$, 
%and the lepton separation near the IP in the $z$ direction to 
we combine a Fisher discriminant 
including 16 modified Fox-Wolfram moments~\cite{sfw}, 
the missing mass $M_{\rm miss} \equiv \sqrt{E_{\rm miss}^{2} - p_{\rm miss}^{2}}$, $\cos\theta_B$, 
and the lepton separation near the IP in the $z$ direction to 
form the likelihood ratio 
%$\mathcal{R}_{B} = {\cal L}_s/({\cal L}_s + {\cal L}_{B\overline{B}})$, 
%where ${\cal L}_{B\overline{B}}$ 
$\mathcal{R}_{\rm sl} = {\cal L}_{\rm s}/({\cal L}_{\rm s} + {\cal L}_{\rm sl})$, 
where $E_{\rm miss}$($p_{\rm miss}$) is the missing energy (momentum), 
$\theta_B$ is the polar angle between the reconstructed $B$ candidate 
and the beam direction in the $\Upsilon(4S)$ rest frame,
and ${\cal L}_{\rm s}$ (${\cal L}_{\rm sl}$) is the likelihood for signal (semileptonic $B$) decays.
%Combinatorial background suppression is improved
%by including $q^2$ and $B$-flavor tagging information~\cite{Kakuno:2004cf}, 
%which is parameterized by a discrete variable $q_{\rm tag}$ indicating 
%the flavor of the tagging $B$-meson candidate and a quality parameter $r$ 
%(ranging from 0 for no flavor information to 1 for unambiguous flavor assignment).
%Selection criteria for $\mathcal{R}$ and $\mathcal{R}_{B}$ are determined by maximizing
%the value of $S/\sqrt{S+B}$, where $S$ and $B$ denote the expected yields 
%of signal and background events in the signal region, respectively,
%in different ($q^2$,\ $q_{\rm rec} \cdot q_{\rm tag} \cdot r$) regions,
%where $q_{\rm rec}$ is the charge of the reconstructed $B$ candidate.
%Events with $q_{\rm rec} \cdot q_{\rm tag} \cdot r$ close to $-1$
%are considered to be well tagged and are unlikely to be from continuum processes.
%For the $K_S^0 \ell^+ \ell^-$ modes, only the dependence on $r$ is considered.
Combinatorial background suppression 
%for both continuum $e^+e^- \to q\overline{q}$ and semileptonic $B$ decay events 
is improved by including $q^2$ and $B$-flavor tagging information~\cite{Kakuno:2004cf}.
Selection criteria for $\mathcal{R}$ and $\mathcal{R}_{\rm sl}$
are determined by maximizing
the value of $S/\sqrt{S+B}$, where $S$ and $B$ denote the expected yields 
of signal and background events in the signal region, respectively, 
in different $q^2$ and tagging regions. 

The dominant backgrounds that peak in the signal region 
are from $B \to J/\psi X$ and $\psi^\prime X$ and 
rejected in the following $q^2$ regions (in units of GeV$^2$/$c^2$): 
%\begin{eqnarray}
%\nonumber 8.68 < &q^2(\mu^+\mu^-)& < 10.09~, \\
%\nonumber 12.86 < &q^2(\mu^+\mu^-)& < 14.18~, \\
%\nonumber 8.11 < &q^2(e^+e^-)& < 10.03~, \\
%\nonumber 12.15 < &q^2(e^+e^-)& < 14.11~. 
%\end{eqnarray}
$8.68 < q^2(\mu^+\mu^-) < 10.09$, $12.86 < q^2(\mu^+\mu^-) < 14.18$, 
$8.11 < q^2(e^+e^-) < 10.03$, and $12.15 < q^2(e^+e^-) < 14.11$.
%The decay $B^+ \to J/\psi(\psi^\prime) h^+$ ($h^+ = K^+,\pi^+$) 
%can also contribute to the $B^+\to K^+ \pi^-\mu^+\mu^-$ and $K_{S}^{0} \pi^+\mu^+\mu^-$ samples 
%if a muon from $J/\psi(\psi^\prime)$ is misidentified as a pion and
%another non-muon track is at the same time misidentified as a muon. 
%We remove such events from the two samples with the requirement 
To remove the background from $B \to J/\psi(\psi^\prime) K^*$ events 
with one of the muons misidentified as a pion candidate,
we reject events with $-0.10$~GeV/$c^2$ $< M_{\pi\mu}-m_{J/\psi(\psi^\prime)} <$ $0.08$~GeV/$c^2$, 
where the pion is assigned the muon mass. 
%The charmonium $B \to D X$ background can contribute to the muon modes 
%if a pion from $D$ meson is misidentified as a muon.
Background from $B \to D X$ is rejected by 
additional veto windows $|M_{K\mu}-m_{D}|<$ 0.02~GeV/$c^2$ and 
%$|M_{K^{*}\mu}-m_{D}|<$ 0.02~GeV/$c^2$.
$|M_{K\pi\mu}-m_{D}|<$ 0.02~GeV/$c^2$, 
where the muon is assigned the pion mass. 
%suppress this background. 
The invariant mass of an electron pair must exceed 
0.14~GeV/$c^2$ in order to remove 
background from photon conversions and $\pi^0 \to \gamma e^+e^-$ decays. 
%In addition, charmless $B \to K^* \pi \pi$ events make a small 
%contribution to the peaking background.

If multiple $B$ candidates survive these selections in an event, 
we select the one with the smallest $|\Delta E|$.
The fractions of multiple $B$ events are about 7\%, 12\%, and 20\% for 
the $K^{+} \pi^{-}$, $K_{S}^{0} \pi^{+}$, and $K^{+} \pi^{0}$ modes, 
respectively, 
and less than 1\% for the $K^+$ and $K^0_S$ modes, according to 
a study using Monte Carlo (MC) simulation of $B \to K^{(*)} \ell \ell$ decays.

To determine the signal yields, 
we perform an extended unbinned maximum likelihood fit 
to $M_{\rm bc}$ and $M_{K\pi}$ for $B \to K^{*} \ell^+ \ell^-$ decays, 
and to $M_{\rm bc}$ for $B \to K \ell^+ \ell^-$ decays. 
%The likelihood function is defined as follows:
%\begin{eqnarray}
%\nonumber
%\mathcal{L} & = & {e^{-(N_s+N_b+N_{c\overline{c} X}+N_{K^{(*)}\pi\pi})}\over N!} \times \\
%\nonumber
%&&\prod_{i=1}^{N}~[
%N_s P_s^{i}+ N_b P_b^{i} + N_{c\overline{c} X} P_{c\overline{c} X}^{i} + 
%N_{K^{(*)}\pi\pi} P_{K^{(*)}\pi\pi}^{i}]~,~~~~~~
%\end{eqnarray}
%where $N$ denotes the number of observed events in the candidate region, 
%and $N_s$ ($P_s^{i}$), $N_b$ ($P_b^{i}$), 
%$N_{c\overline{c} X}$ ($P_{c\overline{c} X}^{i}$), 
%and $N_{K^{(*)}\pi\pi}$ ($P_{K^{(*)}\pi\pi}^{i}$) 
%denote the event yields (the probability density functions, PDFs, for the i-th event)  
%for signal, combinatorial, $B \to J/\psi(\psi^\prime) X$, 
%and $B \to K^{(*)}\pi\pi$ backgrounds. 
The likelihood function includes contributions from signal, combinatorial, 
$B \to J/\psi(\psi^\prime) X$, and $B \to K^{(*)}\pi\pi$ backgrounds.
The signal PDFs consist of a Gaussian (Crystal Ball function~\cite{cbline}) 
in $M_{\rm bc}$ for the muon (electron) modes and 
a relativistic Breit-Wigner shape in $M_{K \pi}$ for the $K^*$ resonance.
The means and widths are determined from MC simulation and calibrated using $B \to J/\psi K^{(*)}$ decays.
The PDFs of signal decays in which either 
the kaon or the pion candidate is wrongly associated to the $K^{*}$ decay (self-cross-feed) 
are modelled by 
%The PDFs of the self-cross-feed, 
%which is from the signal decay with a misreconstructed pion or kaon, 
%are modeled by 
a two-dimensional smoothed histogram function 
with $q^2$-dependent fractions obtained 
from MC simulation of the signal decays.
The combinatorial PDFs are represented by the product of 
an empirical ARGUS function~\cite{argus} in $M_{\rm bc}$ 
with the sum in $M_{K \pi}$ of 
a threshold function (whose threshold is fixed at $m_K + m_{\pi}$) 
and a relativistic Breit-Wigner shape at the $K^*$ resonance. 
The PDFs and yields for $B \to J/\psi(\psi^\prime) X$ decays 
are determined from a large MC sample, 
while the $B \to K^{(*)}\pi\pi$ PDFs and normalizations are determined from data, 
taking into account the probabilities of pions to be misidentified as muons.
Yields for signal and combinatorial background and the combinatorial PDF parameters
are allowed to float in the fit, while the yields and parameters for other components are fixed.

We divide $q^2$ into 6 bins and extract the signal and combinatorial background yields in each bin.
The $K^*$ longitudinal polarization fractions ($F_L$) and $A_{FB}$ are 
extracted from fits to 
$\cos \theta_{K^*}$ and $\cos \theta_{B\ell}$, respectively, 
in the signal region, 
where $\theta_{K^*}$ is the angle between the kaon direction and 
the direction opposite to the $B$ meson 
in the $K^*$ rest frame, 
and $\theta_{B\ell}$ is the angle between the $\ell^+$ ($\ell^-$) 
and the opposite of the $B$ ($\overline{B}$) direction in the dilepton rest frame. 
%The signal PDF for the fit to $\cos \theta_{K^*}$ is described by 
%Considering the spin-0 $B$ and $K$ mesons, 
%the polarized $K^*$ implies an unpolarization in the dilepton system. 
The signal PDFs for the fit to $\cos \theta_{K^*}$ and $\cos \theta_{B\ell}$ are described by 
a product of the $K^*$/dilepton polarization function and the efficiency, 
%as a function of $\cos \theta_{K^*/B\ell}$, 
\begin{eqnarray} 
\nonumber 
%P_{S}(\cos\theta_{K^*})&=&
[~\textstyle{3\over 2} F_L \cos^2 \theta_{K^*} + \textstyle{3\over 4}(1-F_L)(1-\cos^2 \theta_{K^*})~]~
\times \epsilon(\cos \theta_{K^*})~ 
\end{eqnarray}
%where $\epsilon(\cos \theta_{K^*})$ denotes the efficiency obtained from MC. 
%For the fit to $\cos \theta_{B\ell}$, we use 
and
\begin{eqnarray}
\nonumber 
%P_{S}(\cos\theta_{B\ell})&=&
[~\textstyle{3\over 4} F_L (1-\cos^2 \theta_{B\ell}) + \textstyle{3\over 8}(1-F_L)(1+\cos^2 
\theta_{B\ell}) \\
\nonumber 
+ A_{FB}\cos \theta_{B\ell}~] \times \epsilon(\cos \theta_{B\ell})~, 
\end{eqnarray}
respectively. 
%as the signal PDF, 
%where $\epsilon(\cos \theta_{B\ell})$ denotes the efficiency 
%as a function of $\cos \theta_{B\ell}$. 
The first two terms in the dilepton polarization function 
correspond to the production of $K^*$'s with longitudinal and transverse polarization, 
while the third term generates the forward-backward asymmetry.
%the terms in the square brackets are $K^*$ and dilepton polarization functions, 
%and $\epsilon(\cos \theta_{K^*/B\ell})$ denotes the efficiency as a function of 
%$\cos \theta_{K^*/B\ell}$.
%The angular efficiency distributions, background PDFs, and signal and background sizes, 
%obtained from either MC or a $M_{\rm bc}$--$M_{K\pi}$ fit, 
%are fixed in both angular fits. 
%Fig.~\ref{fig:demo_kstllfit} and Fig.~\ref{fig:demo_kllfit} 
%Fig.~\ref{fig:demo_fit} 
Figures in Ref.~\cite{EPAPS} 
illustrate the fits for $B$ yields, $F_L$, and $A_{FB}$ in each $q^2$ bin.
%in one of the 6 $q^2$ bins. 
%(4.30~GeV/$c^2$--8.68~GeV/$c^2$).
In the fit to $\cos \theta_{K^*}$ ($\cos \theta_{B\ell}$), 
$F_L$ ($A_{FB}$) is the only free parameter, 
while the other PDFs and normalizations are fixed.
%in the fit to $\cos \theta_{K^*}$ ($\cos \theta_{B\ell}$).
For the $B\to K \ell^+\ell^-$ modes, 
we set $F_L=1$ and the $B\to K_S^0 \ell^+\ell^-$ sample is not used.

\begin{table*}[htb]
\caption{
Fit results in each of six $q^2$ bins and an additional bin from 1 to 6~GeV$^2$/$c^2$ 
for which recent theory predictions are available~\cite{q2_1-6}. 
The first uncertainties are statistical and the second are systematic.
}
\vskip -0.3cm
\label{tb:q2_result}
\begin{center}
\begin{tabular}{c|ccccc}
\hline
\hline
$q^2$ (GeV$^2$/$c^2$) & $N_s$ & ${\mathcal B}$(10$^{-7}$) & $F_L$ & $A_{FB}$ & $A_I$  \\
\hline
\multicolumn{6}{c}{$B\to K^* \ell^+ \ell^-$} \\
\hline
0.00--2.00      & 27.4$^{+7.4}_{-6.6}$~ & 1.46$^{+0.40}_{-0.35}\pm$0.11~ & 
$0.29^{+0.21}_{-0.18}\pm$0.02~  & 0.47$^{+0.26}_{-0.32}\pm$0.03 & $-0.67^{+0.18}_{-0.16}\pm$0.05~ \\
2.00--4.30      & 16.8$^{+6.1}_{-5.3}$~ & 0.86$^{+0.31}_{-0.27}\pm$0.07~ & 
$0.71^{+0.24}_{-0.24}\pm$0.05~  & 0.11$^{+0.31}_{-0.36}\pm$0.07 & $1.45^{+1.04}_{-1.15}\pm$0.10~ \\
4.30--8.68    & 27.9$^{+9.5}_{-8.5}$~ & 1.37$^{+0.47}_{-0.42}\pm$0.39~ & 
$0.64^{+0.23}_{-0.24}\pm$0.07~  & 0.45$^{+0.15}_{-0.21}\pm$0.15 & $-0.34^{+0.29}_{-0.27}\pm$0.14~ \\
10.09--12.86 & 54.0$^{+10.5}_{-9.6}$~& 2.24$^{+0.44}_{-0.40}\pm$0.19~ & 
$0.17^{+0.17}_{-0.15}\pm$0.03~  & 0.43$^{+0.18}_{-0.20}\pm$0.03 &  $0.00^{+0.20}_{-0.21}\pm$0.09~ \\
14.18--16.00    & 36.2$^{+9.9}_{-8.8}$~ & 1.05$^{+0.29}_{-0.26}\pm$0.08~ & 
$-0.15^{+0.27}_{-0.23}\pm$0.07~ & 0.70$^{+0.16}_{-0.22}\pm$0.10 &  $0.16^{+0.30}_{-0.35}\pm$0.09~ \\
$>$16.00       & 84.4$^{+11.0}_{-9.9}$~& 2.04$^{+0.27}_{-0.24}\pm$0.16~ & 
0.12$^{+0.15}_{-0.13}\pm$0.02~  & 0.66$^{+0.11}_{-0.16}\pm$0.04 & $-0.02^{+0.20}_{-0.21}\pm$0.09~ \\
\hline
1.00--6.00      & 29.42$^{+8.9}_{-8.0}$~ & 1.49$^{+0.45}_{-0.40}\pm$0.12~ & 
$0.67^{+0.23}_{-0.23}\pm$0.05~  & 0.26$^{+0.27}_{-0.30}\pm$0.07 &  $0.33^{+0.37}_{-0.43}\pm$0.08~ \\
\hline
\multicolumn{6}{c}{$B\to K \ell^+ \ell^-$} \\
\hline
%$q^2$ (GeV$^2$/$c^2$) & $N_s$ & ${\mathcal B}$(10$^{-7}$) & $A_I$  \\
0.00--2.00      & 27.0$^{+6.0}_{-5.4}$~ & 0.81$^{+0.18}_{-0.16}\pm$0.05~  
&$-$&  $0.06^{+0.32}_{-0.35}\pm$0.02 & $-0.33^{+0.33}_{-0.25}\pm$0.08 \\
2.00--4.30      & 17.6$^{+5.5}_{-4.8}$~ & 0.46$^{+0.14}_{-0.12}\pm$0.03~  
&$-$& $-0.43^{+0.38}_{-0.40}\pm$0.09 & $-0.47^{+0.50}_{-0.38}\pm$0.07 \\
4.30--8.68      & 39.1$^{+7.5}_{-6.9}$~ & 1.00$^{+0.19}_{-0.18}\pm$0.06~  
&$-$& $-0.20^{+0.12}_{-0.14}\pm$0.03 & $-0.19^{+0.25}_{-0.21}\pm$0.08 \\
10.09--12.86    & 22.0$^{+6.2}_{-5.5}$~ & 0.55$^{+0.16}_{-0.14}\pm$0.03~ 
&$-$& $-0.21^{+0.17}_{-0.15}\pm$0.06 & $-0.29^{+0.37}_{-0.29}\pm$0.08 \\
14.18--16.00    & 15.6$^{+4.9}_{-4.3}$~ & 0.38$^{+0.19}_{-0.12}\pm$0.02~  
&$-$&  $0.04^{+0.32}_{-0.26}\pm$0.05 & $-0.40^{+0.61}_{-0.69}\pm$0.07 \\
$>$16.00       & 40.3$^{+8.2}_{-7.5}$~ & 0.98$^{+0.20}_{-0.18}\pm$0.06~  
&$-$& $0.02^{+0.11}_{-0.08}\pm$0.02  &  $0.11^{+0.24}_{-0.21}\pm$0.08 \\
\hline
1.00--6.00      & 52.0$^{+8.7}_{-8.0}$~ & 1.36$^{+0.23}_{-0.21}\pm$0.08~   
&$-$& $-0.04^{+0.13}_{-0.16}\pm$0.05 & $-0.41^{+0.25}_{-0.20}\pm$0.07 \\
\hline
\hline
\end{tabular}
\end{center}
%\vskip -0.8cm
\end{table*} 

\begin{figure*}[htb]
\begin{center}
%\vskip -1.0cm \hskip -6cm
%\includegraphics[width=6cm,height=5.5cm]{dbf_kstll.eps}\\
%\vskip -5.5cm \hskip 6cm
%\includegraphics[width=6cm,height=4cm]{ai.eps}\\
%\vskip -0.7cm \hskip 6cm
%\includegraphics[width=6cm,height=4cm]{fl.eps}\\
%\vskip -2.3cm \hskip -6cm
%\includegraphics[width=6cm,height=5.5cm]{dbf_kll.eps}\\
%\vskip -3.9cm \hskip 6cm
%\includegraphics[width=6cm,height=4cm]{afb.eps}\\
\includegraphics[height=8cm]{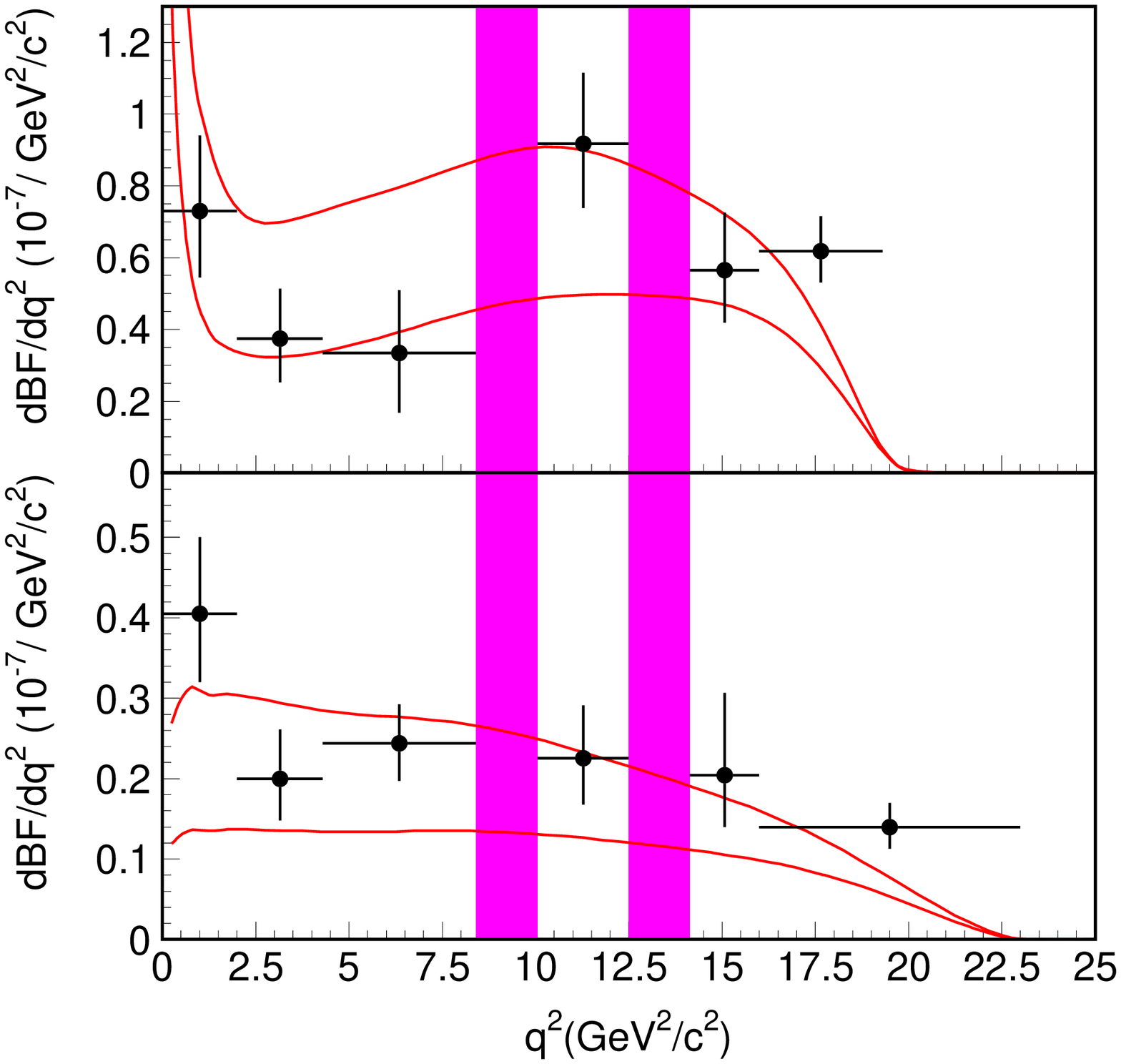}
\includegraphics[height=8cm]{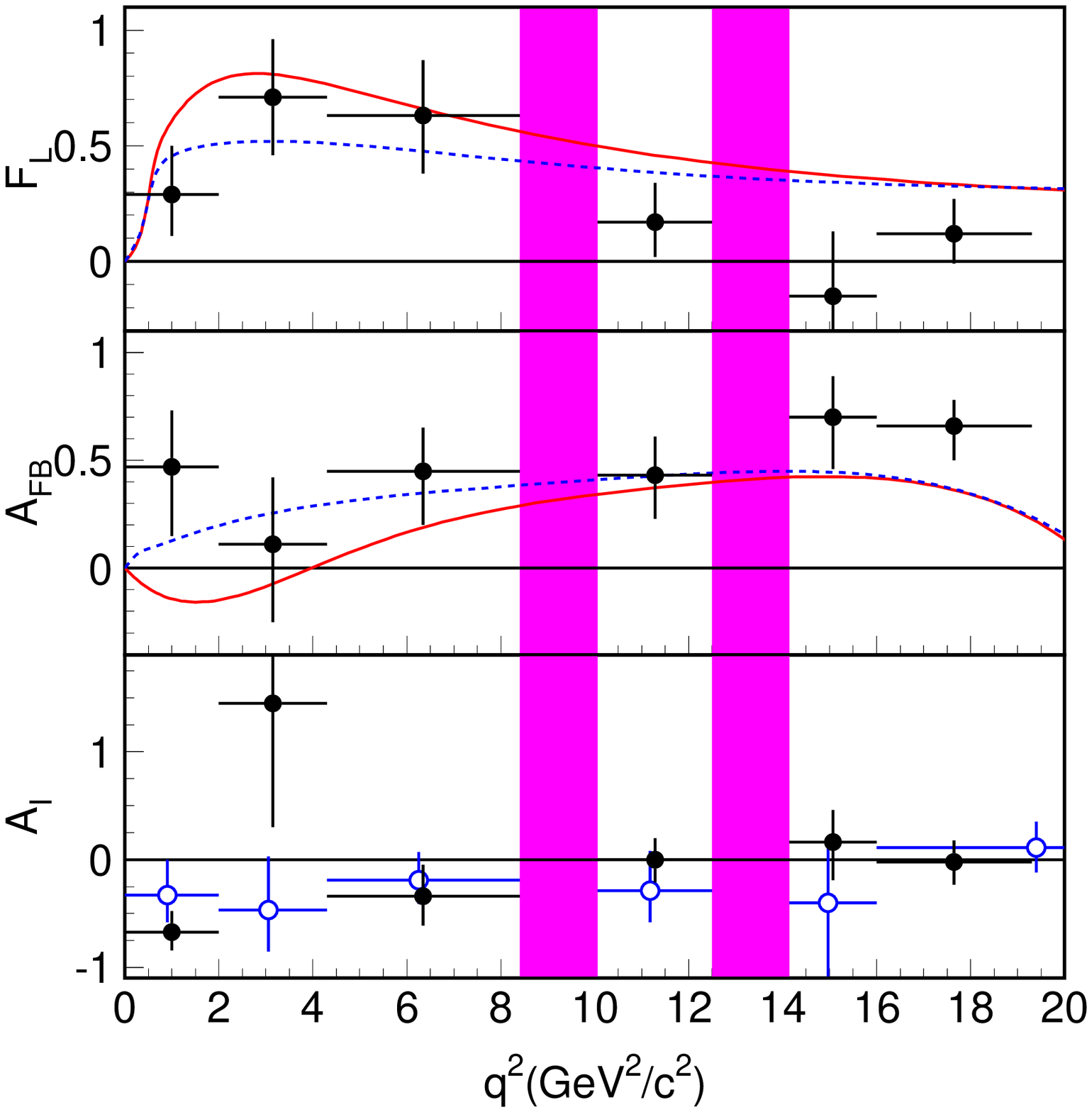}\\
%\vskip -7.4cm \hskip -9.5cm {\bf (a)} \\
%\vskip -0.2cm \hskip 2.5cm {\bf (c)} \\
%\vskip 1.7cm \hskip 2.5cm {\bf (d)} \\
%\vskip 0.7cm \hskip -9.5cm {\bf (b)} \\
%\vskip 0.6cm \hskip 2.5cm {\bf (e)} \\
\vskip -7.4cm \hskip -2.5cm {\bf (a)} \\
\vskip -0.4cm \hskip 14cm {\bf (c)} \\
\vskip 1.5cm \hskip 14cm {\bf (d)} \\
\vskip 0.5cm \hskip -2.5cm {\bf (b)} \\
\vskip 0.5cm \hskip 14cm {\bf (e)} \\
\vskip 2.4cm
\end{center}
\caption{
Differential branching fractions for the 
(a) $K^* \ell^+ \ell^-$ and (b) $K \ell^+ \ell^-$ modes as a function of $q^2$. 
The two shaded regions are veto windows to reject $J/\psi(\psi^\prime) X$ events. 
%The solid curves are the theoretical prediction~\cite{Ali:01}.
The solid curves show the SM theoretical predictions 
with the minimum and maximum allowed form factors~\cite{Ali:01}.
(c) and (d) show the fit results for $F_L$ and $A_{FB}$ 
in $K^* \ell^+ \ell^-$ as a function of $q^2$, 
together with the solid (dotted) curve representing the SM ($C_7=-C^{SM}_7$) prediction~\cite{Ali:01}.
(e) is the $A_I$ asymmetry as a function of $q^2$ for the 
$K^* \ell^+ \ell^-$ (filled circles) and $K \ell^+ \ell^-$ (open circles) modes.
%The theoretical curves for $F_L$ and $A_{FB}$ spectrums are not sensitive to 
%the form factor-related uncertainties.
}
\vskip -0.4cm
\label{fig:result}
\end{figure*}

Table~\ref{tb:q2_result} lists the measurements of $B$ yields; 
%A table in Ref.~\cite{EPAPS} lists the measurements of $B$ yields; 
the partial branching fractions, 
obtained by correcting the $B$ yields for $q^2$-dependent efficiencies; 
$F_L$; and $A_{FB}$ in individual $q^2$ bins.
%With the assumption of the total branching fraction ratios for the electron mode to the muon mode 
%being 1.33 for $B\to K^* \ell^+ \ell^-$ and 1.0 for $B\to K \ell^+ \ell^-$, 
%different ratios are treated for different $q^2$ bins 
%based on the theoretical prediction~\cite{Ali:01}. 
In the calculation of the partial branching fractions, 
we adopt the SM lepton flavor ratios of the muon to electron modes~\cite{Ali:01} and 
express the branching fractions in terms of the muon channel.
The ratio for the full $q^2$ interval is 
$R_{K^*}^{\rm SM}=0.75$ ($R_{K}^{\rm SM}=1$)
for the $B\to K^* \ell^+ \ell^-$ ($B\to K \ell^+ \ell^-$) mode, 
where the deviation from unity is due to the photon pole.
%The differential branching fraction, $F_L$, and $A_{FB}$ as functions of $q^2$ 
%for the $K^{*} \ell^+ \ell^-$ and $K \ell^+ \ell^-$ modes 
The results, as well as the SM curves, are shown in Fig.~\ref{fig:result}. 
To illustrate how non-SM physics might manifest itself, 
we superimpose curves on the $F_L$ and $A_{FB}$ plots corresponding to the case of 
%flipped $C_7$ ($C_7=-C^{SM}_7$), 
$C_7$ with reversed sign ($C_7=-C^{SM}_7$). 
The measured values do not reject this possibility.
%The sign of $C_7$ is neither well-understood theoretically nor constrained experimentally. 
 
The total branching fractions, 
extrapolated from the partial branching fractions, 
are measured to be~\cite{error}
\begin{eqnarray}
\nonumber
\mathcal{B}(B \to K^* \ell^+ \ell^-) & = & (10.7^{+1.1}_{-1.0}\pm0.9) \times 10^{-7}~,\\
\nonumber
\mathcal{B}(B \to K \ell^+ \ell^-) & = & (4.8^{+0.5}_{-0.4}\pm0.3) \times 10^{-7}~; 
\end{eqnarray}
while the fitted $CP$ asymmetries, 
%defined as $A_{CP} \equiv (N_{\overline{B}} - N_{B})/ (N_{\overline{B}} + N_{B})$, 
%where $N_B (N_{\overline{B}})$ stands for the $B$ ($\overline{B}$) yield, 
defined in terms of the $\overline{B}$ ($B$) yield $N_b$ $(N_{\bar{b}})$ 
as $A_{CP} \equiv (N_{b} - N_{\bar{b}})/ (N_{b} + N_{\bar{b}})$, 
are~\cite{EPAPS} 
\begin{eqnarray}
\nonumber
A_{CP}(K^{*} \ell^+ \ell^-) & = & -0.10\pm0.10\pm0.01~,\\
\nonumber
A_{CP}(K^+ \ell^+ \ell^-) & = & 0.04\pm0.10\pm0.02~. 
\end{eqnarray}
%We calculate the ratios of branching fractions for the electron mode to the muon mode.

The lepton flavor ratio 
%for $B\to K^* \ell^+ \ell^-$ ($R_{K^*}$) is 
%sensitive to the size of the photon pole and is predicted to be {0.75} in the SM, 
%while the ratio for $B\to K \ell^+ \ell^-$ ($R_K$) 
is sensitive to Higgs emission 
and is predicted to be larger than the SM value in the Higgs doublet model at 
large $\tan\beta$~\cite{susy}.
The measured ratios are 
\begin{eqnarray}
\nonumber
R_{K^*} & = & 0.83\pm0.17\pm0.08~,\\
\nonumber
R_{K} & = & 1.03\pm0.19\pm0.06~.
\end{eqnarray}

%The isospin asymmetry, shown in the table~\cite{EPAPS} and Fig.~\ref{fig:result}, 
The isospin asymmetry, shown in Table~\ref{tb:q2_result} and Fig.~\ref{fig:result}, 
is defined as 
\begin{eqnarray}
\nonumber
A_I \equiv \frac{(\tau_{B^+}/\tau_{B^0}) \times {\mathcal B}(K^{(*)0} \ell^+ \ell^-) - {\mathcal B}(K^{(*)\pm} \ell^+ \ell^-)}
{(\tau_{B^+}/\tau_{B^0}) \times {\mathcal B}(K^{(*)0} \ell^+ \ell^-) + {\mathcal B}(K^{(*)\pm} \ell^+ \ell^-)}~,
\end{eqnarray}
where $\tau_{B^+}/\tau_{B^0} = 1.071$ is the lifetime ratio of $B^+$ to $B^0$~\cite{pdg}.
A large isospin asymmetry for $q^2$ below the mass of the $J/\psi$ resonance was reported recently~\cite{ai_exp}. 
We also measure the combined $A_I$ for $q^2<$ 8.68~GeV$^2$/$c^2$ and find  
\begin{eqnarray}
\nonumber
A_I(B \to K^* \ell^+ \ell^-) & = -0.29^{+0.16}_{-0.16} \pm0.09~& ~\sigma = 1.37~,\\
\nonumber
A_I(B \to K \ell^+ \ell^-) & = -0.31^{+0.17}_{-0.14} \pm0.08~& ~\sigma = 1.75~,\\
\nonumber
A_I(B \to K^{(*)} \ell^+ \ell^-) & = -0.30^{+0.12}_{-0.11} \pm0.08~& ~\sigma = 2.22~,
\end{eqnarray}
where $\sigma$ denotes the significance from a null asymmetry and is defined as 
$\sigma \equiv \sqrt{-\rm{2ln} \left( \mathcal{L}_{0} / \mathcal{L}_{\rm max} \right)}$, 
where $\mathcal{L}_{0}$ is the likelihood with $A_I$ constrained to be zero 
and $\mathcal{L}_{\rm max}$ is the maximum likelihood. 
Systematic uncertainties are considered in the significance calculation.
%No significant isospin asymmetry but same sign as the Ref.~\cite{ai_exp} is found 
%in the low $q^2$ region.
No significant isospin asymmetry is found at low $q^2$.

Systematic uncertainties in the branching fraction measurements arise predominantly from 
tracking efficiencies (2.0\%--4.4\%), MC decay models (0.9\%--4.6\%), 
electron (3.0\%) and muon (2.6\%) identification, 
$K^0_S$ (4.9\%) and $\pi^0$ (4.0\%) reconstruction, 
and $\mathcal{R}$ and $\mathcal{R}_{\rm sl}$ selection (1.2\%--3.6\%).
The MC simulated samples of the signal 
are generated based on a model derived from Ref.~\cite{Ali:01}. 
The modeling uncertainties are evaluated by comparing MC samples 
based on different models~\cite{model}, 
while lepton identification is studied using a $J/\psi \to \ell^+ \ell^-$ data sample. 
For $\mathcal{R}$ and $\mathcal{R}_{\rm sl}$ selections, 
we estimate the uncertainties from large control samples with the same final states, 
$B \to J/\psi K^{(*)}$ with $J/\psi \to \ell^+ \ell^-$.
Other uncertainties such as kaon and pion identification efficiencies, 
fitting PDFs, background contamination from $J/\psi$ decays and charmless $B$ decays, 
and the number of $B\overline{B}$ pairs are found to be small.
%Other uncertainties such as kaon and pion identification efficiencies (1.0\%), 
%fitting PDFs (1.4\%--4.6\%), 
%background contamination from $J/\psi$ decays and charmless $B$ decays (0.3\%--7.8\%), 
%and number of $B\overline{B}$ pairs (1.4\%) are found to be small.
The total systematic uncertainties in the branching fractions 
for different decay channels are 
6.8\%--12.2\% and 5.2\%--7.4\% for the $K^{*} \ell^+ \ell^-$ and 
$K \ell^+ \ell^-$ modes, respectively.

The main uncertainties for angular fits  
are propagated from the errors on the fixed normalizations and $F_L$, 
determined from $M_{\rm bc}$--$M_{K\pi}$ and $\cos \theta_{K^*}$ fits, respectively. 
Fitting bias and fitting PDFs are checked using large $B \to J/\psi K^{(*)}$ and MC samples. 
The total uncertainties for the $F_L$ and $A_{FB}$ fits depend on the $q^2$ bin
and range from 0.02--0.06 and 0.03--0.13, respectively. 
The systematic errors on $A_{CP}$ 
are assigned using the $CP$ asymmetry measured in sideband data 
without $\mathcal{R}$ and $\mathcal{R}_{\rm sl}$ selections and are found to be 0.01--0.02.
The systematic error on $R_{K^{(*)}}$ ($A_I$)  
is determined by combining the uncertainties from lepton ($K$/$\pi$) identification, 
$\mathcal{R}$ and $\mathcal{R}_{\rm sl}$ selections, fitting PDFs and background contamination. 
The uncertainty in $A_I$ from the assumption 
of equal production of $B^0\bar{B^0}$ and $B^+B^-$ pairs is also considered. 
The correlated systematic errors among $q^2$ bins are negligible 
for all the measurements. 

In summary, 
we report the differential branching fraction, isospin asymmetry, 
$K^*$ longitudinal polarization and forward-backward asymmetry as functions of $q^2$, 
as well as total branching fractions, lepton flavor ratios, and $CP$ asymmetries 
for $B\to K^{(*)} \ell^+ \ell^-$.
%in both $B\to K^* \ell^+ \ell^-$ and $B\to K \ell^+ \ell^-$ decays.
%$K^*$ longitudinal polarization and forward-backward asymmetry 
%as functions of $q^2$ in $B \to K^* \ell^+ \ell^-$ 
%are also measured from an angular analysis. 
%These results supersede the previous and recent BaBar 
%measurements~\cite{ai_exp,kll_exp} with better accuracy.
These results supersede our previous measurements~\cite{kll_exp_bl} and
are consistent with the latest BaBar results~\cite{ai_exp,kll_exp_bb} with better precision.
The differential branching fraction, lepton flavor ratios, and $K^*$ polarization are 
%in good agreement 
consistent with the SM predictions.
%Neither significant $CP$ nor isospin asymmetry is found.
%However, the later shows the same sign as the BaBar result~\cite{ai_exp}. 
No significant $CP$ asymmetry is found in the study. 
%The isospin asymmetry fluctuates around zero in the bins at low $q^2$; however, 
The isospin asymmetry does not deviate significantly from the null value.
The $A_{FB}(q^2)$ spectrum for $B\to K^* \ell^+ \ell^-$ decays
%although consistent with previous measurements~\cite{kll_exp}, 
tends to be shifted toward the positive side from the SM expectation.
The cumulative difference between the SM 
prediction and the measured points is found to be 2.7 standard deviations. 
%especially at large $q^2$ values. 
%For $B\to K \ell^+ \ell^-$, it is found to be consistent with zero as expected. 
A much larger data set, 
which will be available at the proposed
super $B$ factory~\cite{superB} and LHCb~\cite{LHCb}, 
is needed to make more precise tests of the SM.
%and other theoretical scenarios.

% Please paste this acknowledgement into your latex file. 
% updated 12/15/08   add Nagoya's TLPRC, 2 Grant-in-Aids (long only)
%                        2 new NNSFC contract no. (long only) 
% updated 11/26/08   Poland: KBN -> MNiSW, Australia: DEST -> DISR
%
%***** Acknowledgments *****
%-------- Short version, if necessary, for PRL -----------
We thank the KEKB group for excellent operation of the
accelerator, the KEK cryogenics group for efficient solenoid
operations, and the KEK computer group and
the NII for valuable computing and SINET3 network support.  
We acknowledge support from MEXT, JSPS and Nagoya's TLPRC (Japan);
ARC and DIISR (Australia); NSFC (China); 
DST (India); MOEHRD and KOSEF (Korea); MNiSW (Poland); 
MES and RFAAE (Russia); ARRS (Slovenia); SNSF (Switzerland); 
NSC and MOE (Taiwan); and DOE (USA).

%-- History of updates (for record)
% updated 9/23/08,9/5/08   Koera; new KOSEF grant no. and KRF removed
% updated 2/5/08     Super SINET -> SINET3
% corrected 10/26/07 China missed "Natural" is recovered (Long only)
% updated 9/18/07    China, KIP of CAS removed, new cont. no.
% updated 1/23/07    add RFAE and change MIST -> MES for Russia
% updated 11/22/06   Chinese Academy of Sciencies -> Sciences
% updated 6/23/06
% updated 2/21/06, 1/28/06, 12/25/05
% short version reduced 8/11/05
% updated 7/17/05, 2/17/05

\newpage
\appendix
APPENDIX\\

Below is EPAPS supplementary material for fit figures, 
as well as the total branching fractions 
and $CP$ asymmetries for individual $K^{(*)}\ell^+\ell^-$ modes, 
which is mentioned in Ref.~\cite{EPAPS}.
 
%Fig.~\ref{fig:kstllfit} and Fig.~\ref{fig:kllfit} illustrate the fit results; 
%while the Table~\ref{tb:tbf} shows the total branching fractions and $CP$ asymmetries 
%for individual $K^{(*)} \ell^+ \ell^-$ modes. 
%They are regarded as EPAPS materials and cited in the paper submitted to journal.

\begin{figure}[htb]
\begin{center}
\vskip -0.5cm 
\includegraphics[width=7.5cm]{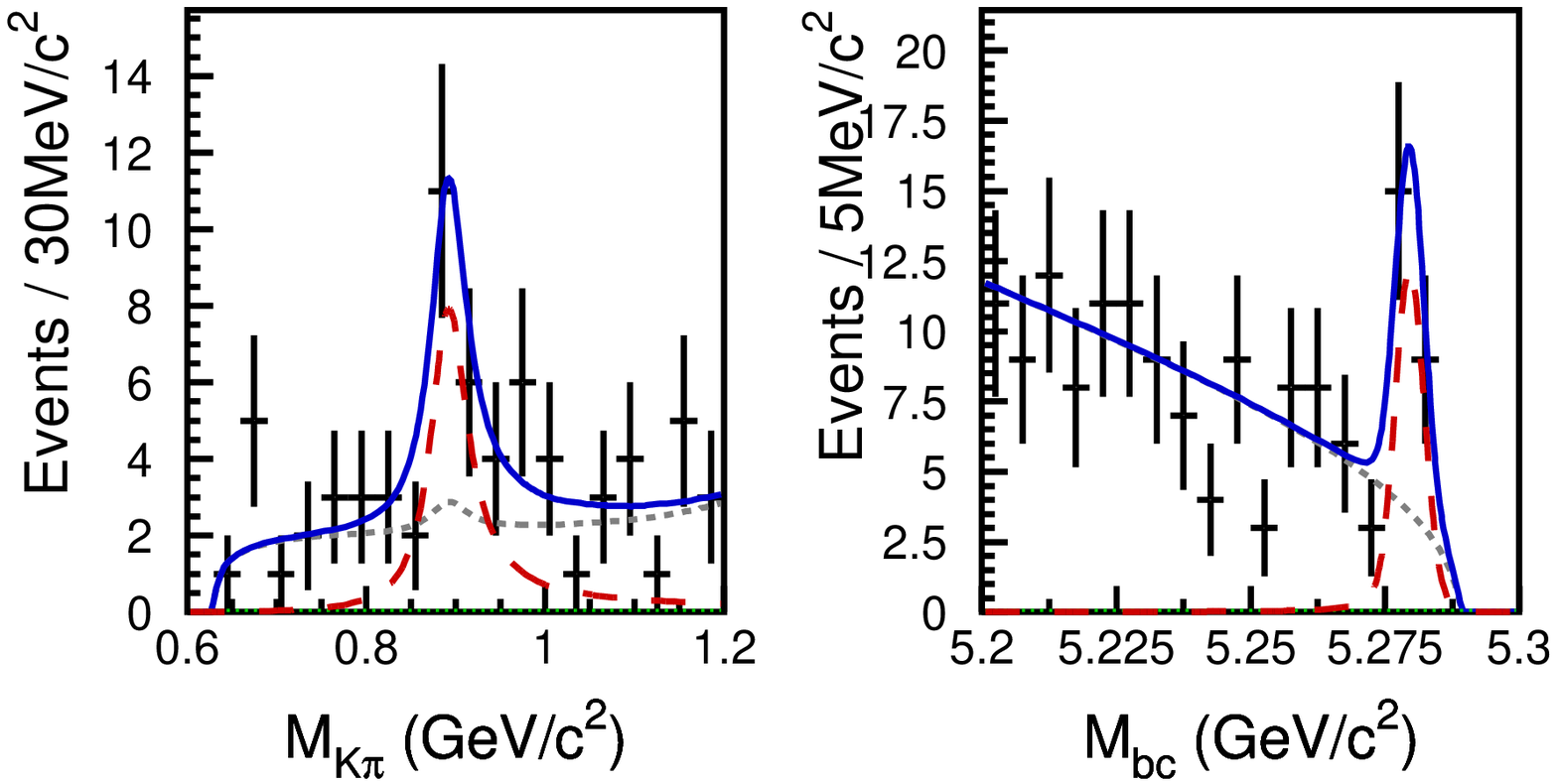} 
\hskip -0.5cm
\includegraphics[width=4.5cm,height=3.5cm]{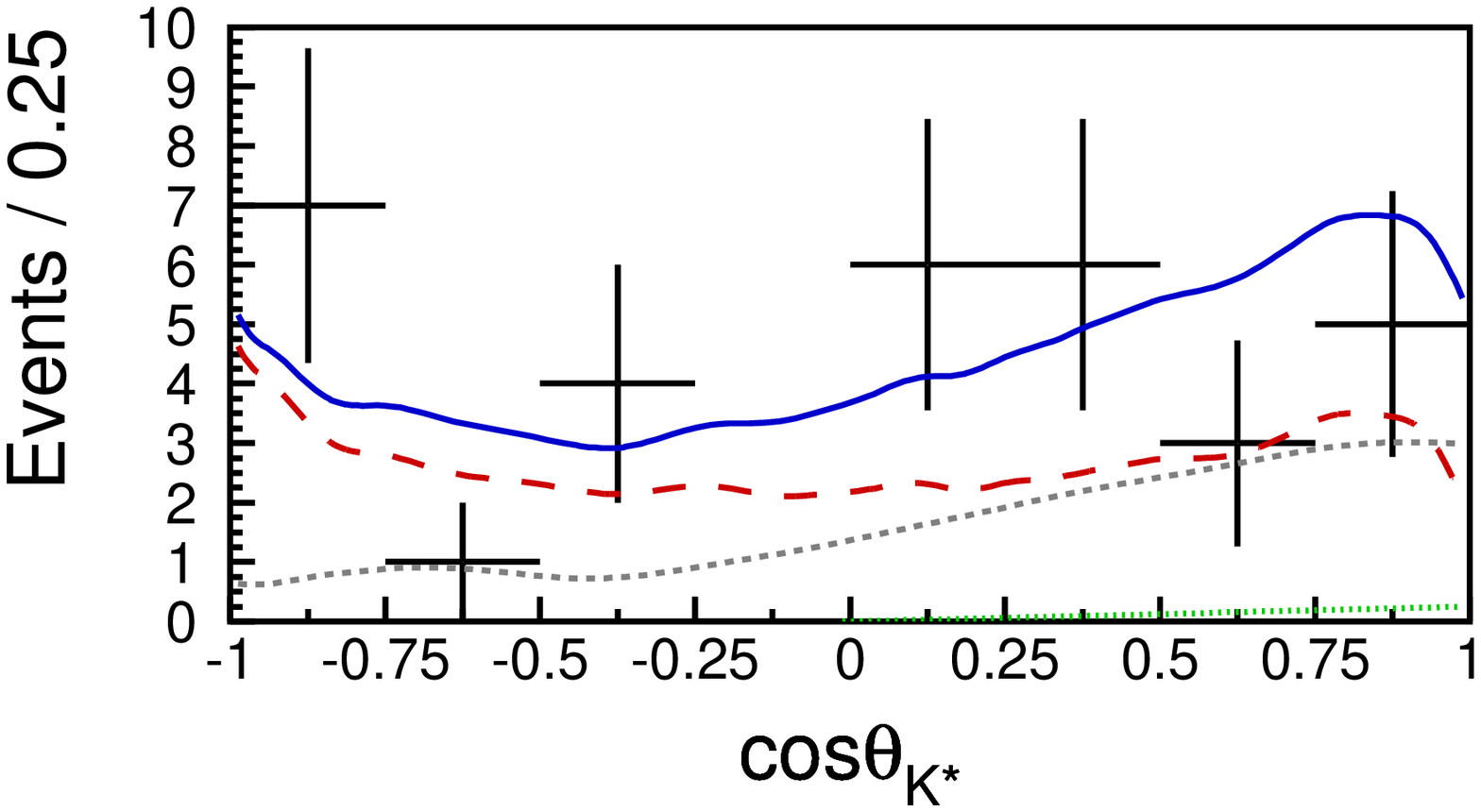} 
\hskip -0.5cm
\includegraphics[width=4.5cm,height=3.5cm]{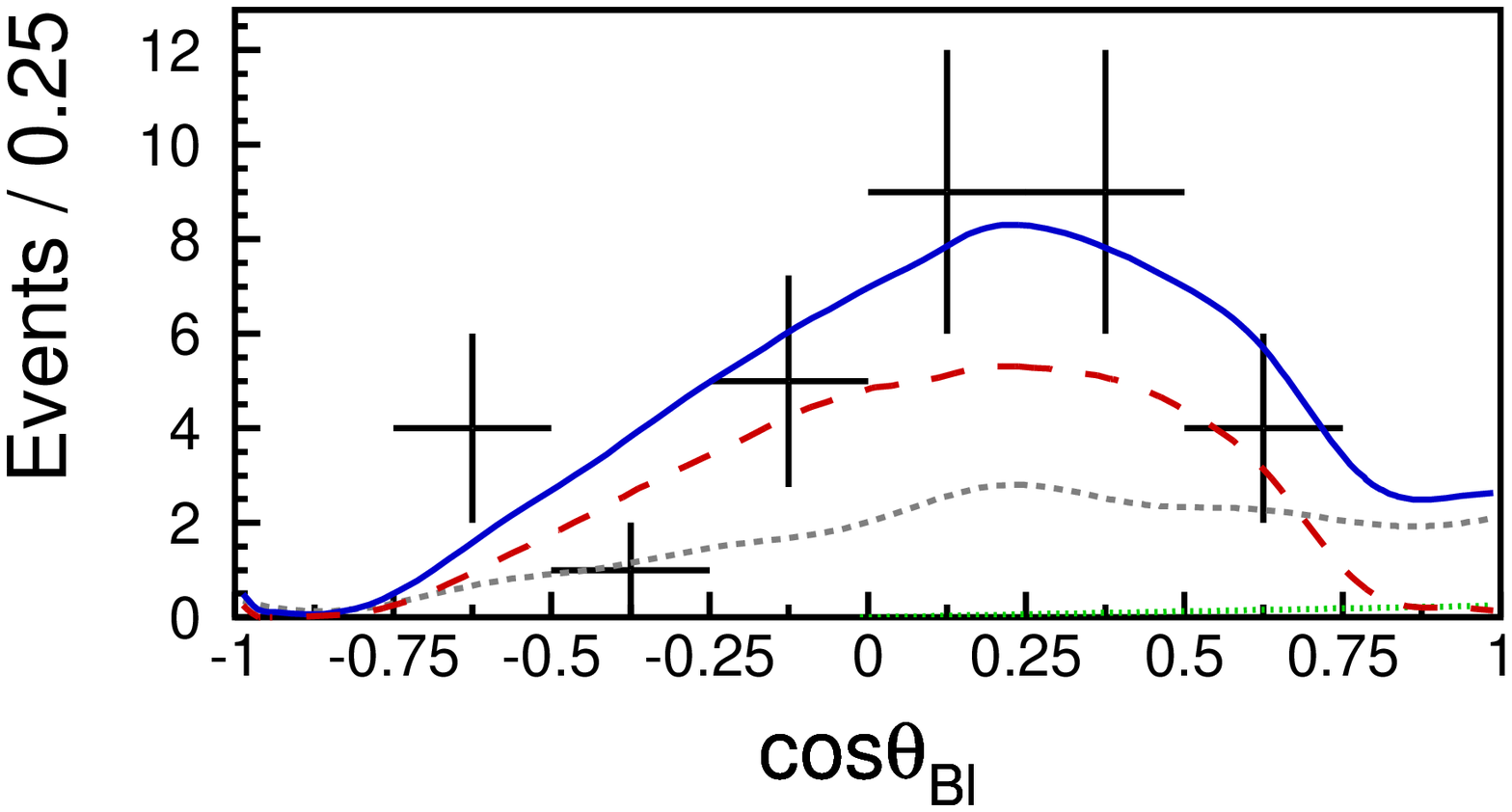} \\
\vskip -3.6cm \hskip -6.4cm {\bf (a)} \hskip 2.7cm {\bf (b)} \\
\vskip -0.1cm \hskip 10.2cm {\bf (c)} \hskip 3.4cm {\bf (d)} \\
\vskip 2.2cm
\includegraphics[width=7.5cm]{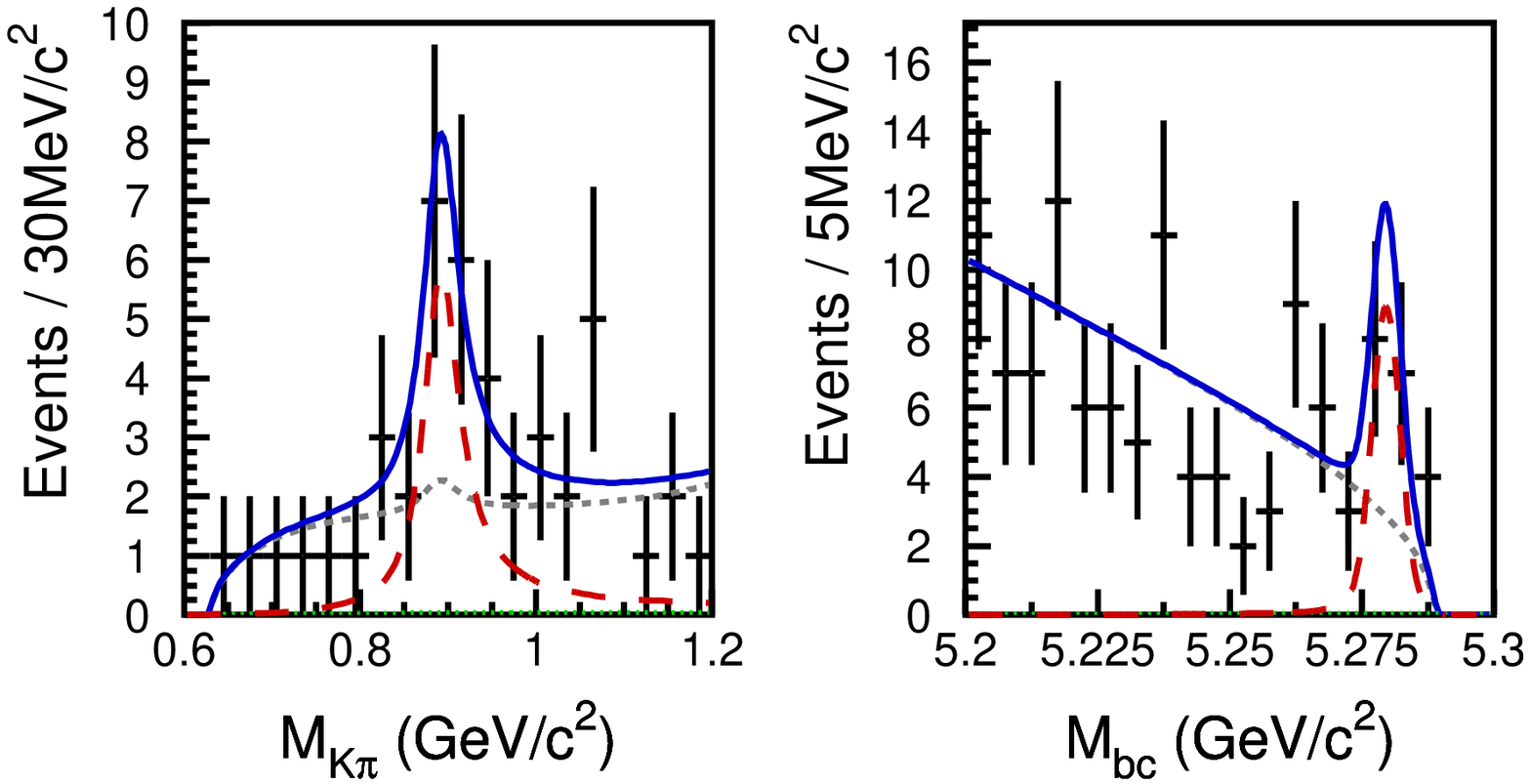} 
\hskip -0.5cm
\includegraphics[width=4.5cm,height=3.5cm]{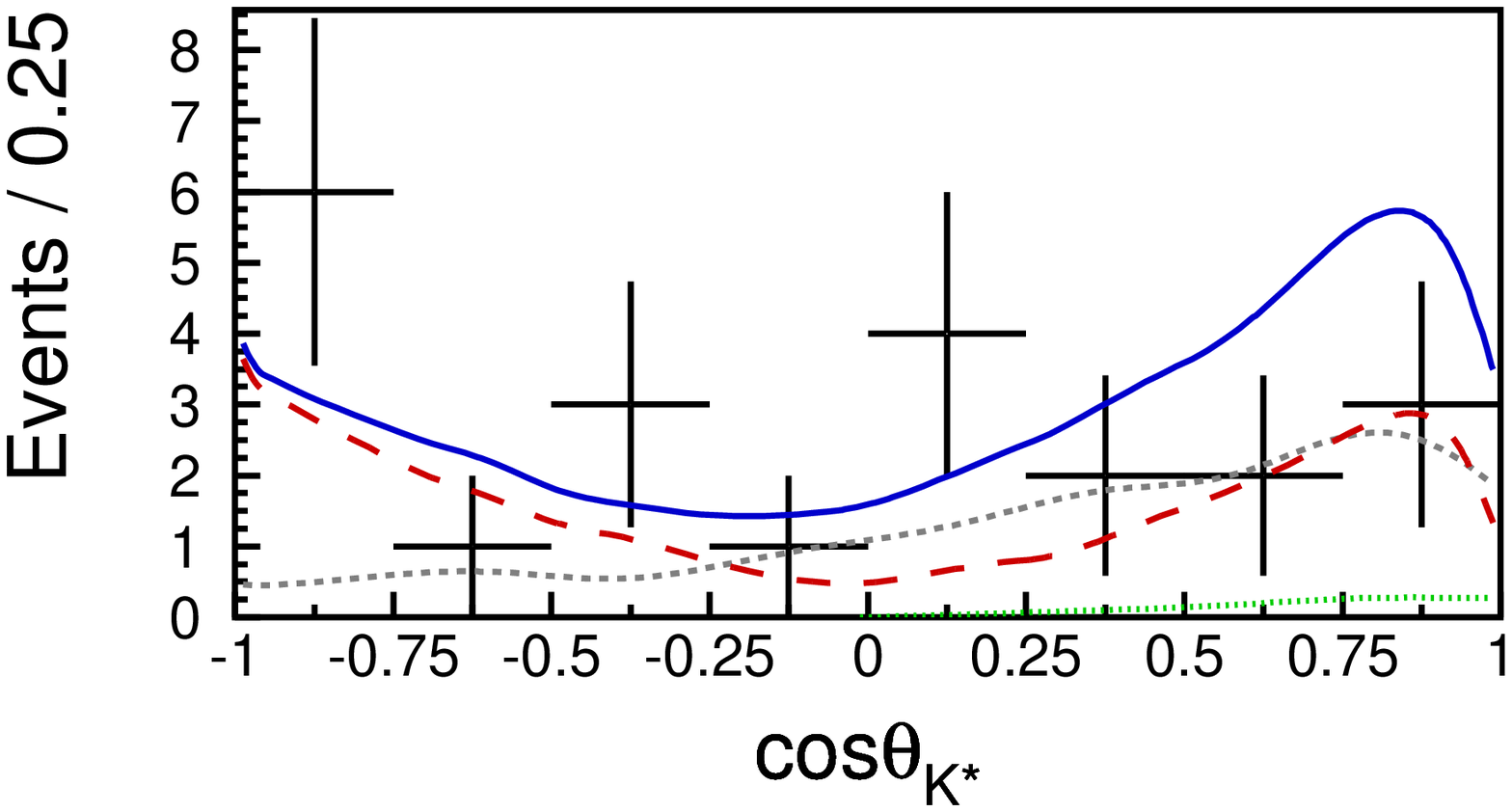} 
\hskip -0.5cm
\includegraphics[width=4.5cm,height=3.5cm]{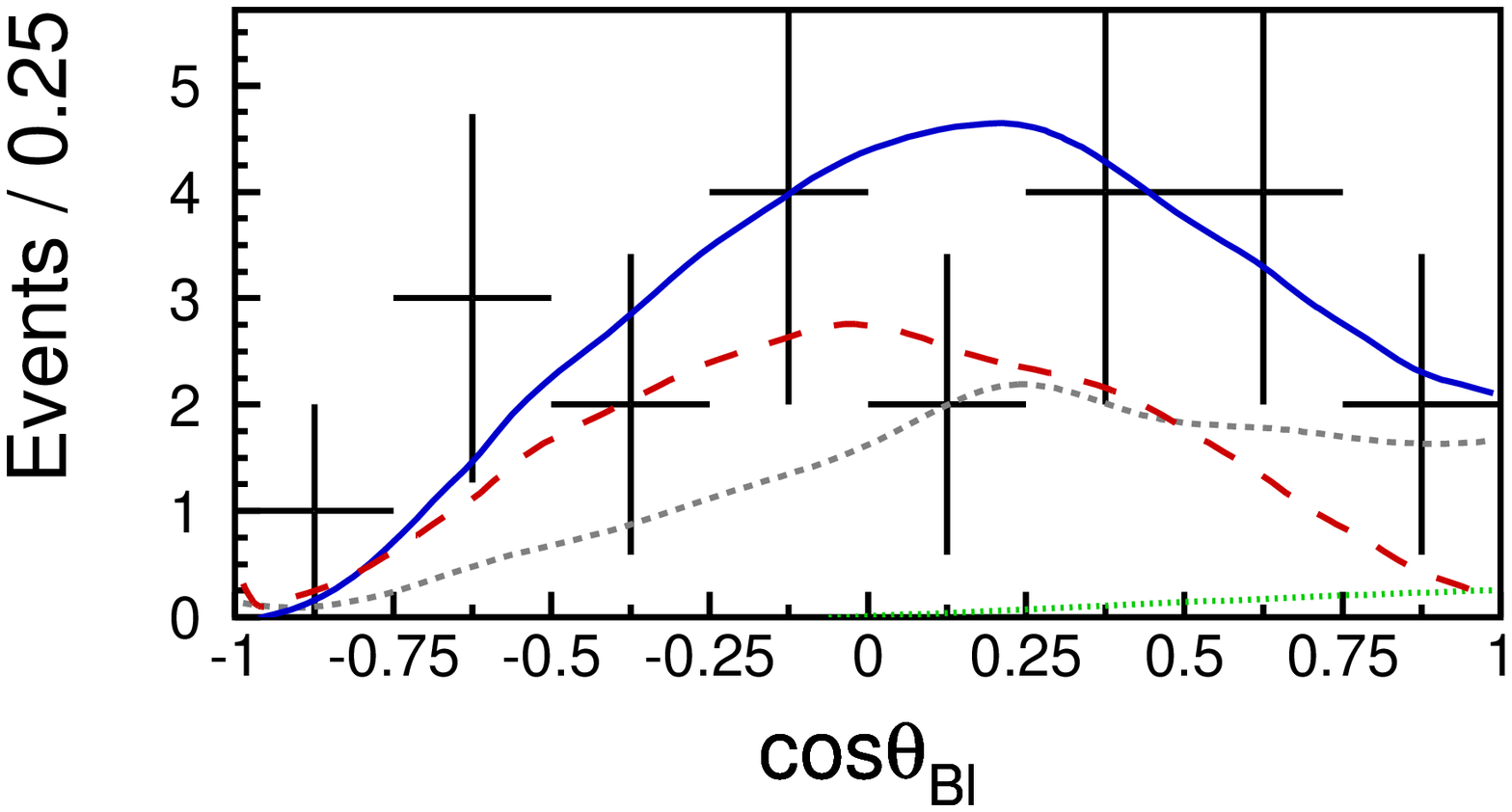} \\
\vskip -3.6cm \hskip -6.4cm {\bf (e)} \hskip 2.7cm {\bf (f)} \\
\vskip -0.1cm \hskip 10.2cm {\bf (g)} \hskip 3.4cm {\bf (h)} \\
\vskip 2.2cm
\includegraphics[width=7.5cm]{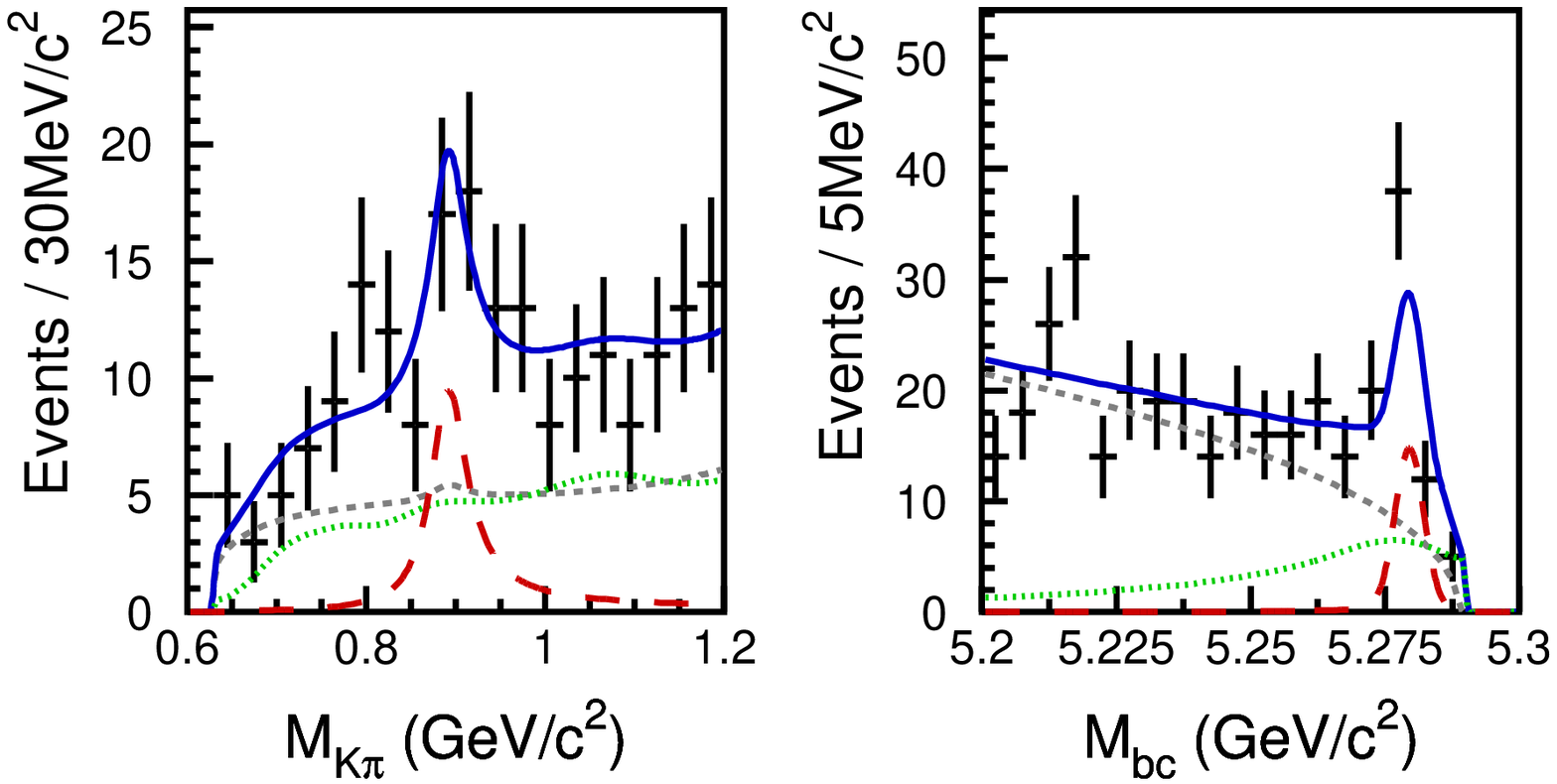} 
\hskip -0.5cm
\includegraphics[width=4.5cm,height=3.5cm]{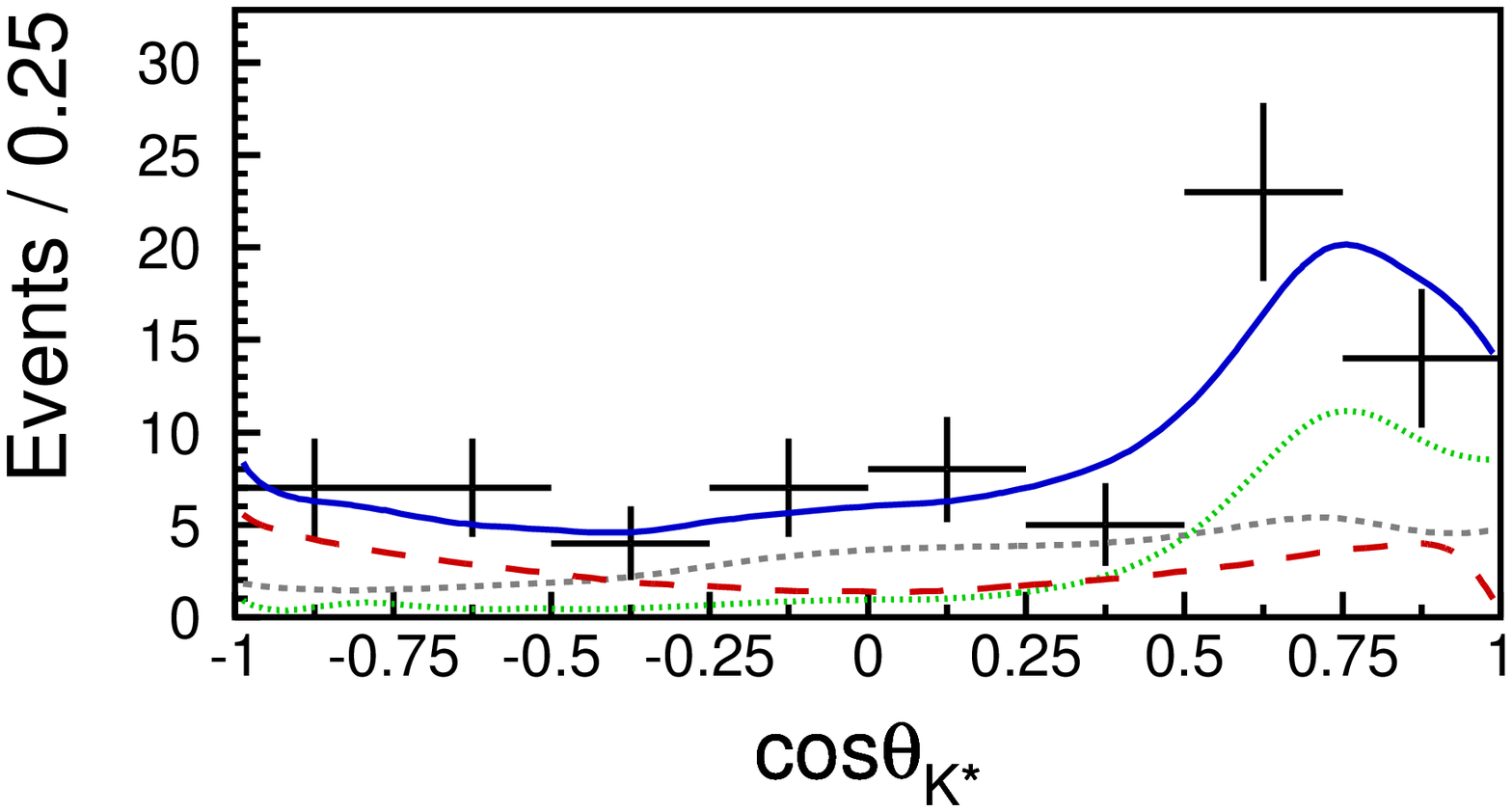} 
\hskip -0.5cm
\includegraphics[width=4.5cm,height=3.5cm]{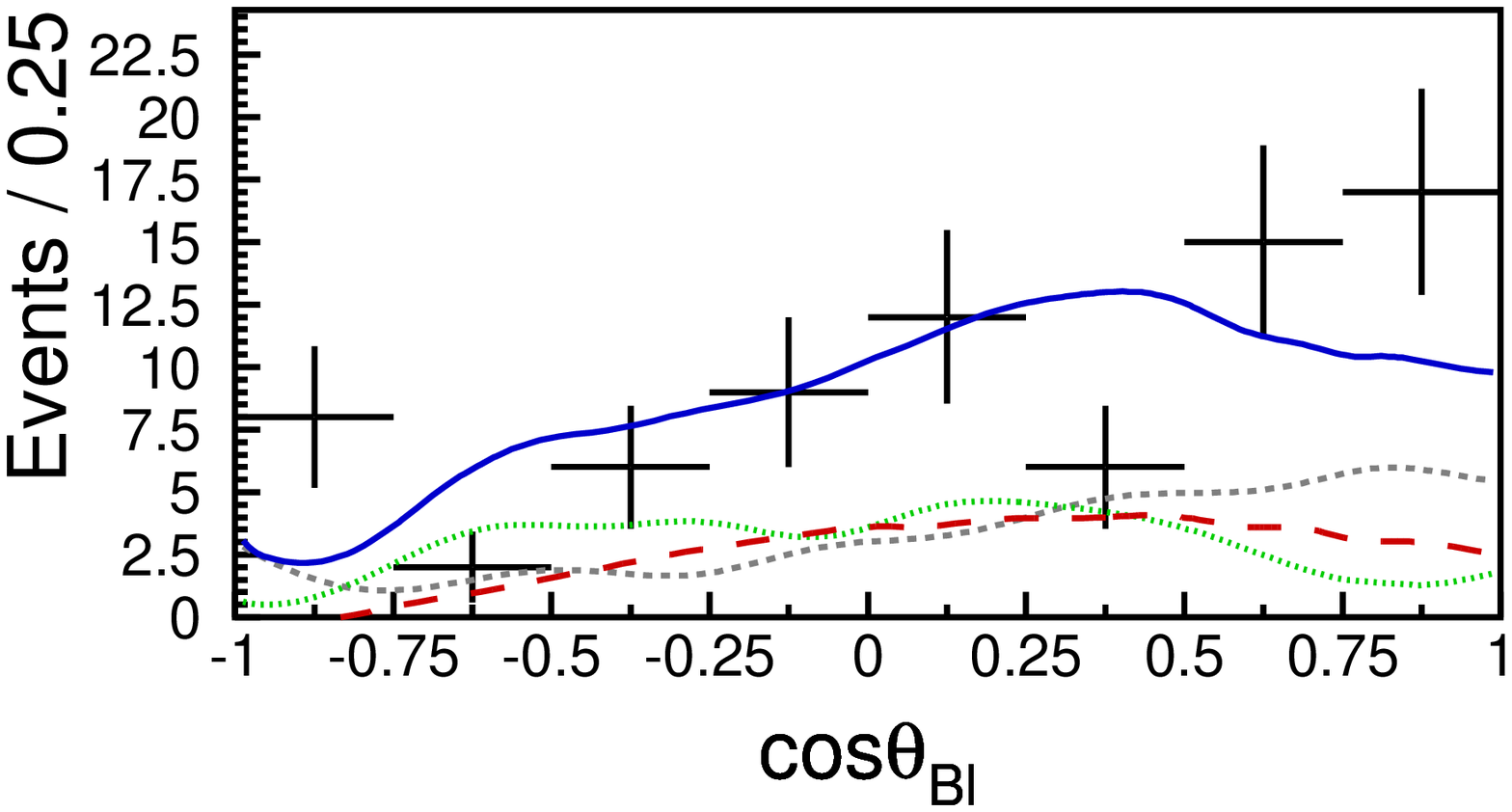} \\
\vskip -3.6cm \hskip -6.4cm {\bf (i)} \hskip 2.7cm {\bf (j)} \\
\vskip -0.1cm \hskip 10.2cm {\bf (k)} \hskip 3.4cm {\bf (l)} \\
\vskip 2.2cm
\includegraphics[width=7.5cm]{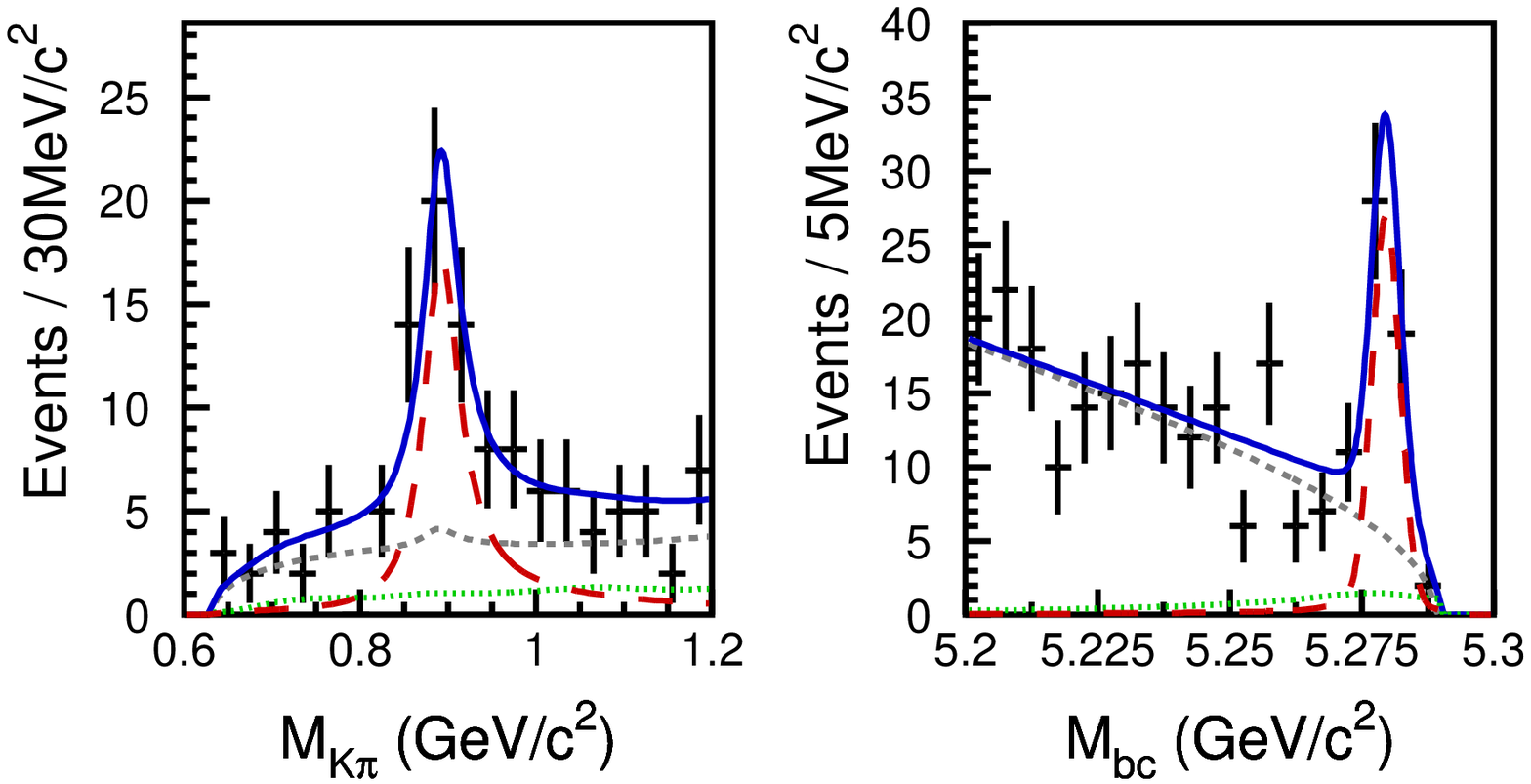} 
\hskip -0.5cm
\includegraphics[width=4.5cm,height=3.5cm]{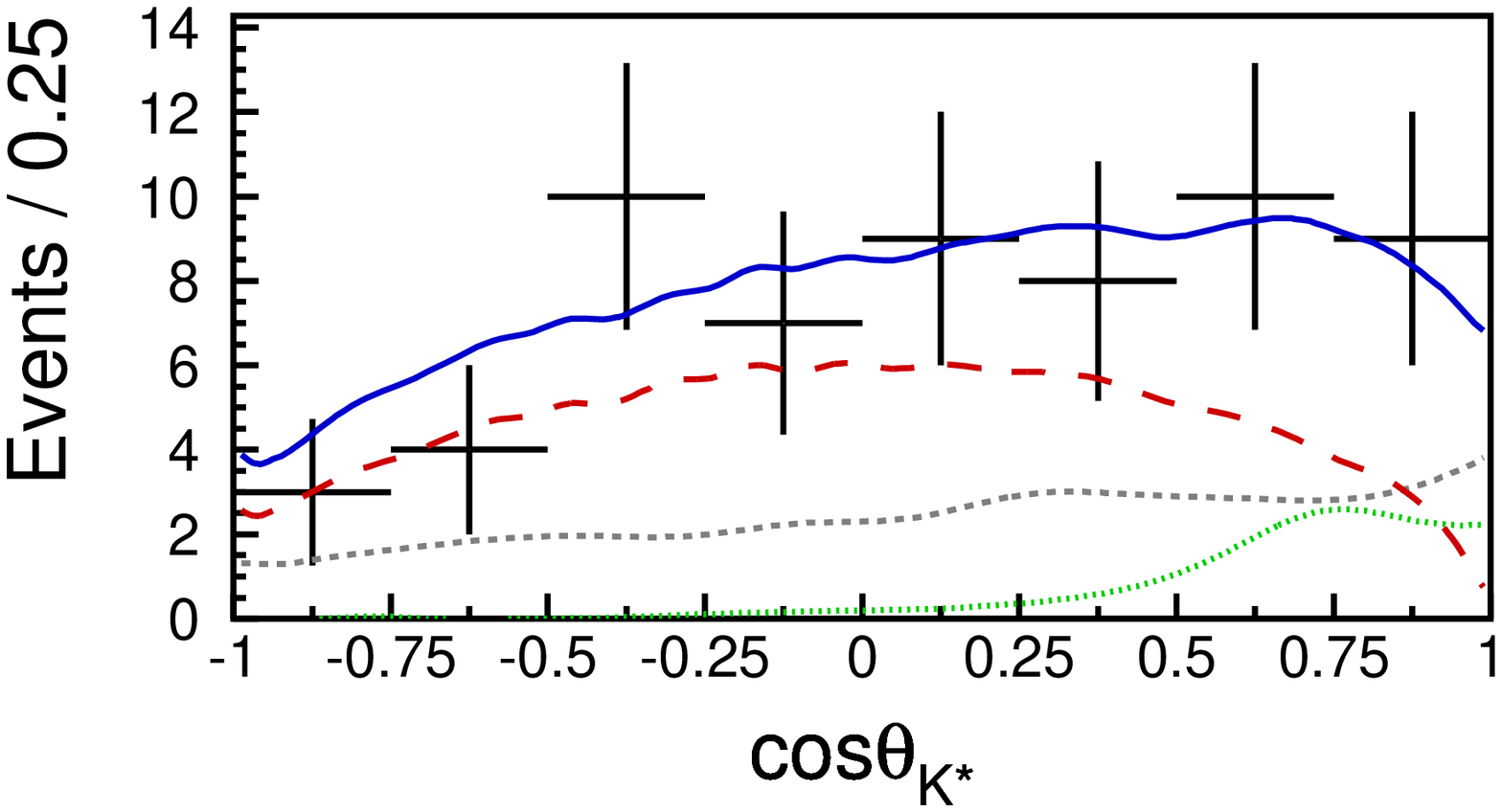} 
\hskip -0.5cm
\includegraphics[width=4.5cm,height=3.5cm]{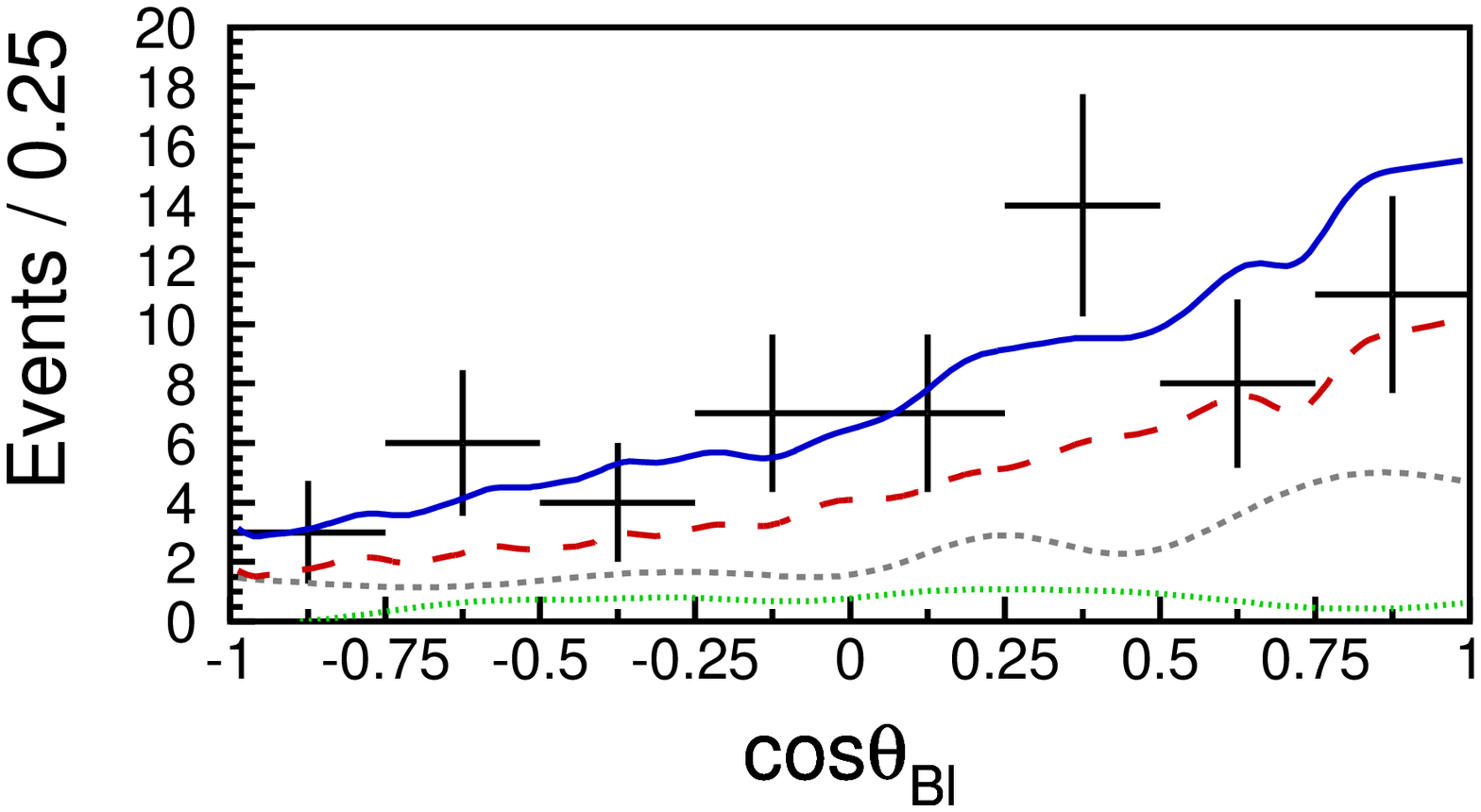} \\
\vskip -3.6cm \hskip -6.4cm {\bf (m)} \hskip 2.7cm {\bf (n)} \\
\vskip -0.1cm \hskip 10.2cm {\bf (o)} \hskip 3.4cm {\bf (p)} \\
\vskip 2.2cm
\includegraphics[width=7.5cm]{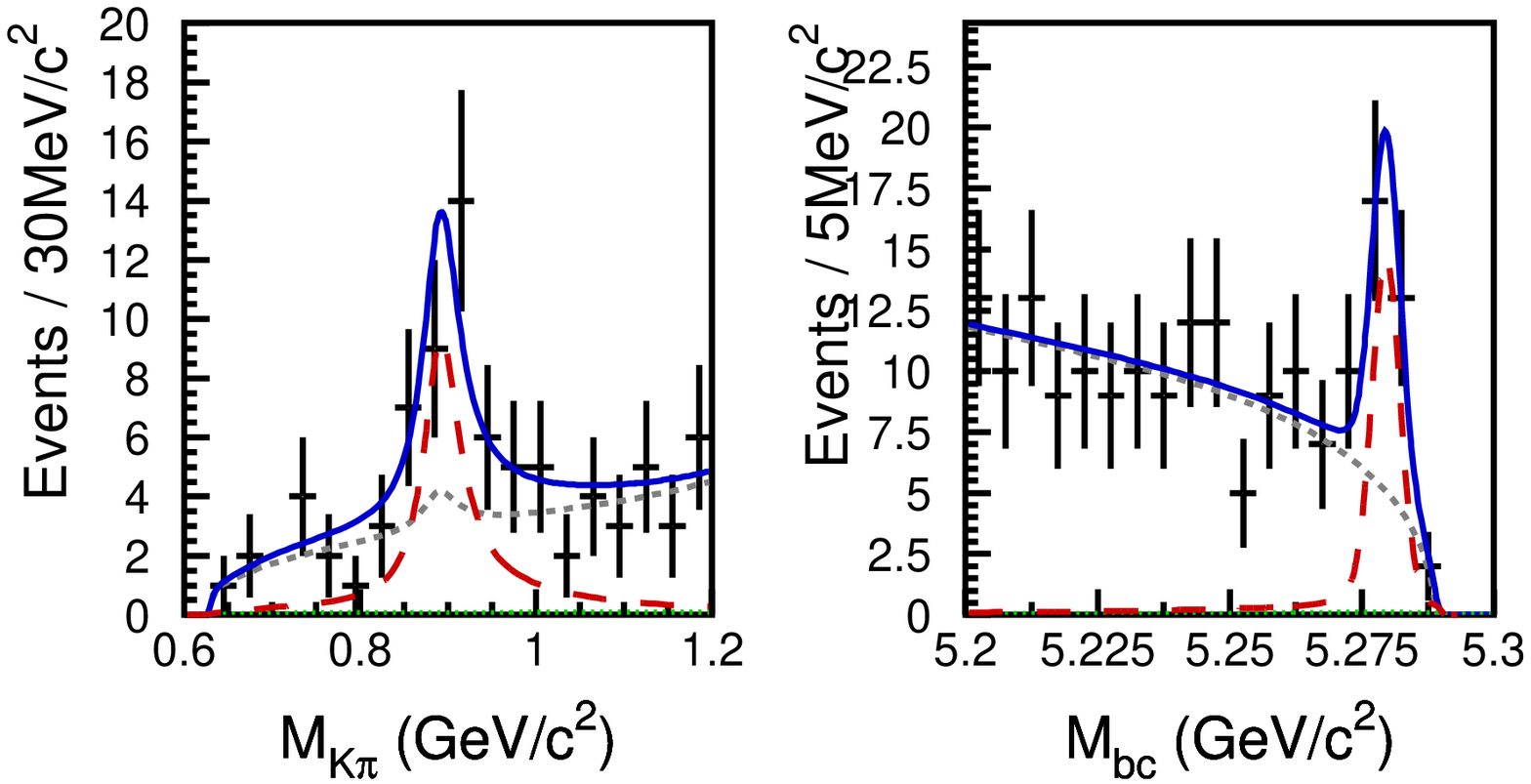} 
\hskip -0.5cm
\includegraphics[width=4.5cm,height=3.5cm]{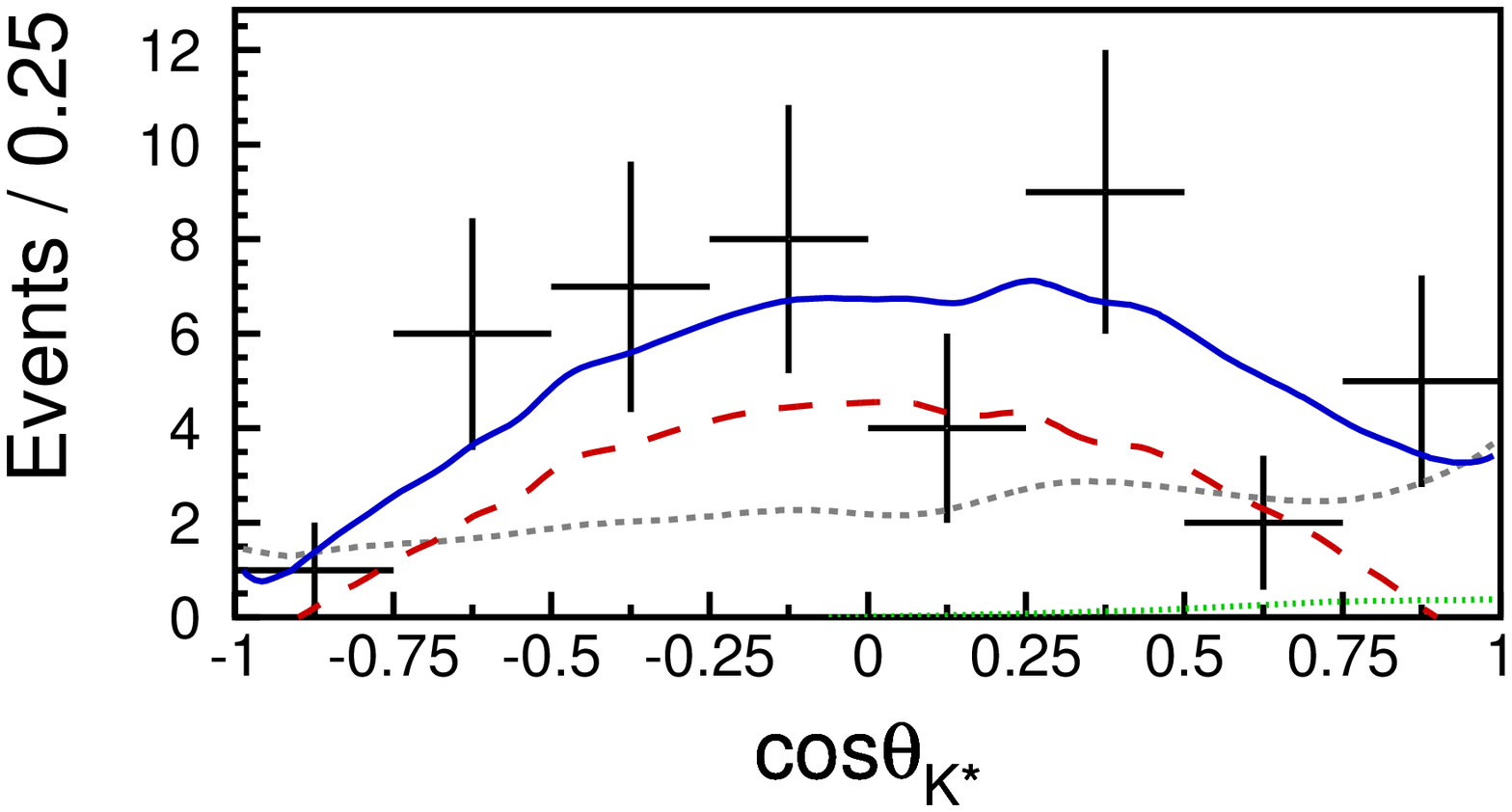} 
\hskip -0.5cm
\includegraphics[width=4.5cm,height=3.5cm]{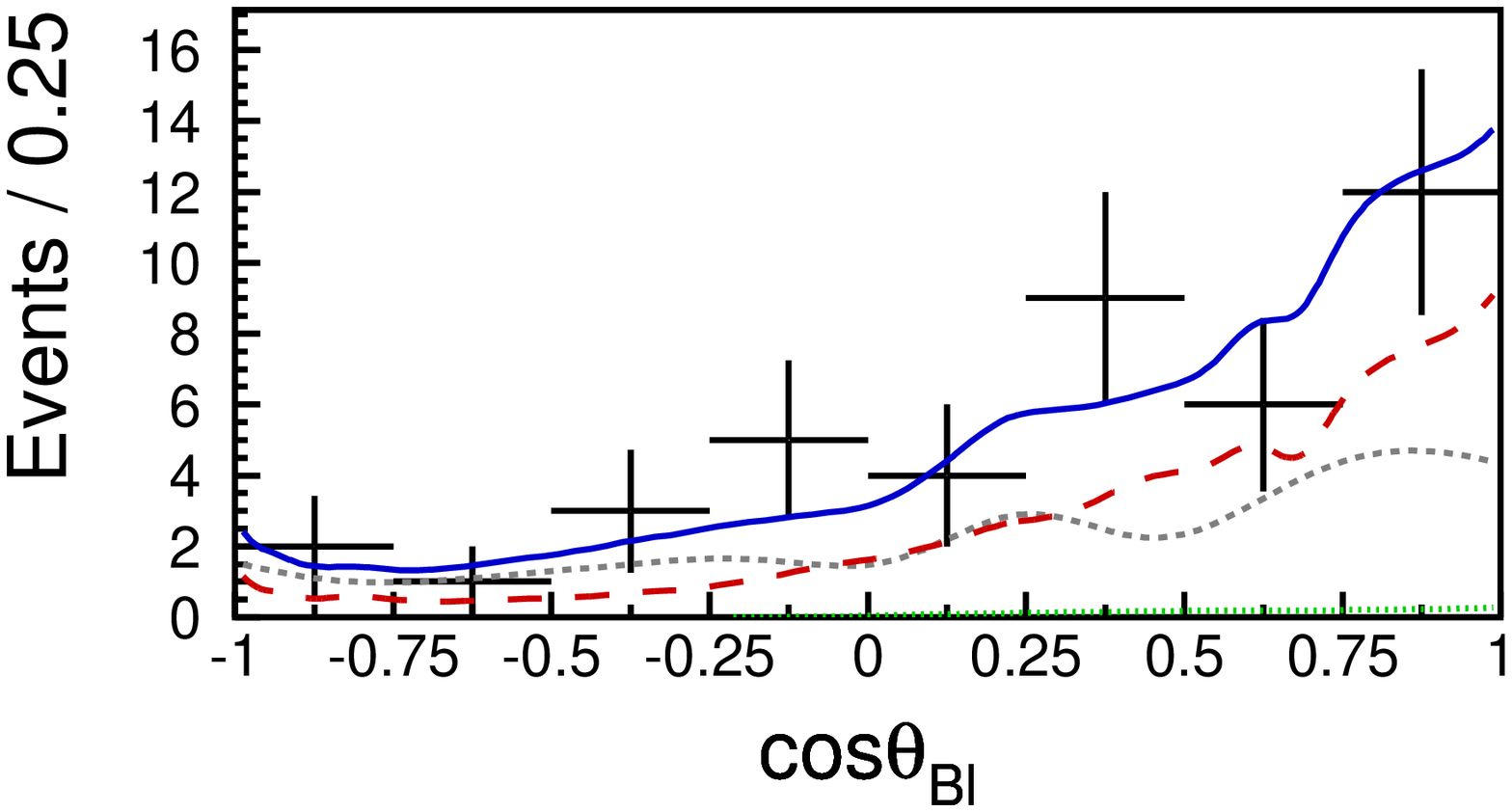} \\
\vskip -3.6cm \hskip -6.4cm {\bf (q)} \hskip 2.7cm {\bf (r)} \\
\vskip -0.1cm \hskip 10.2cm {\bf (s)} \hskip 3.4cm {\bf (t)} \\
\vskip 2.2cm
\includegraphics[width=7.5cm]{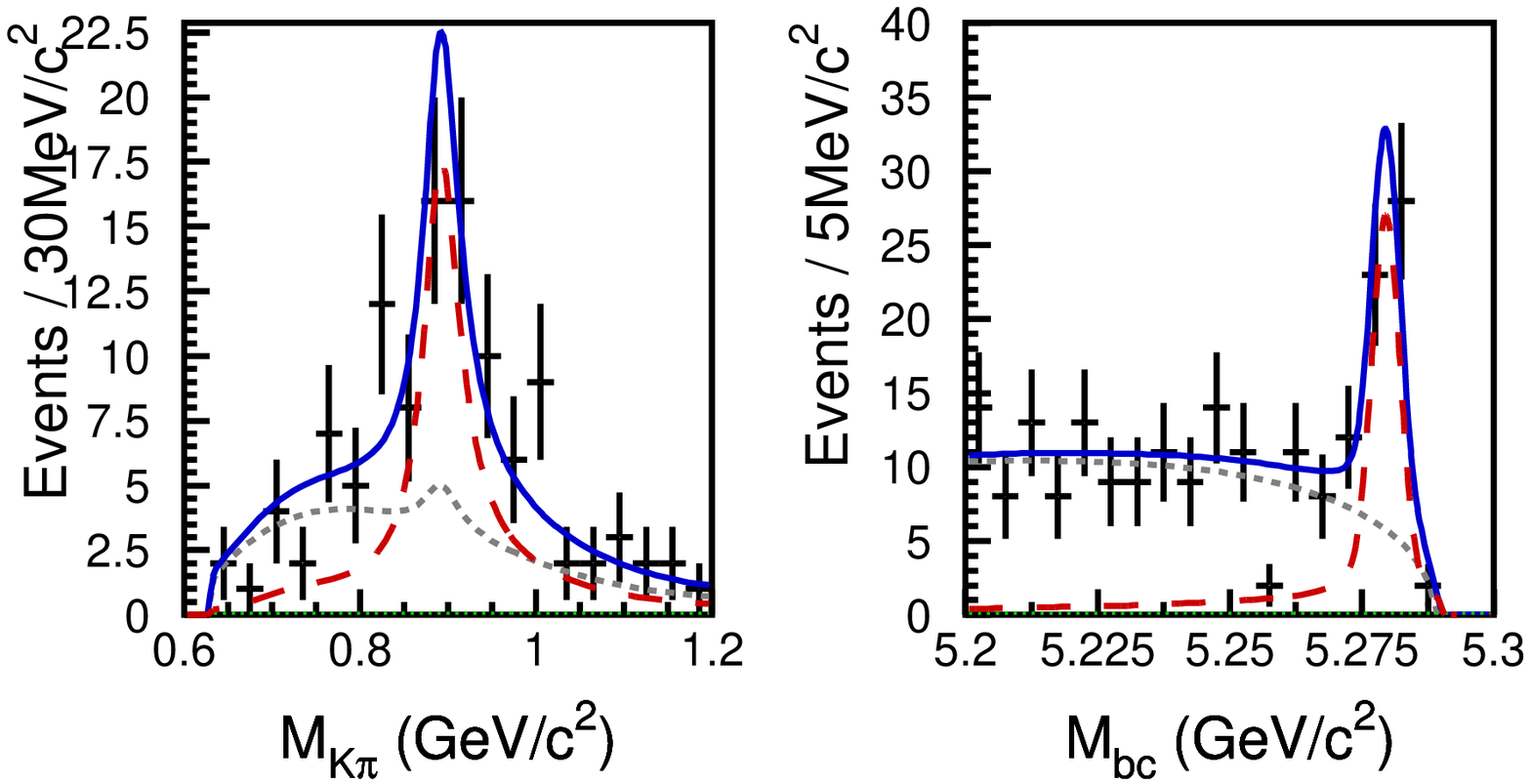} 
\hskip -0.5cm
\includegraphics[width=4.5cm,height=3.5cm]{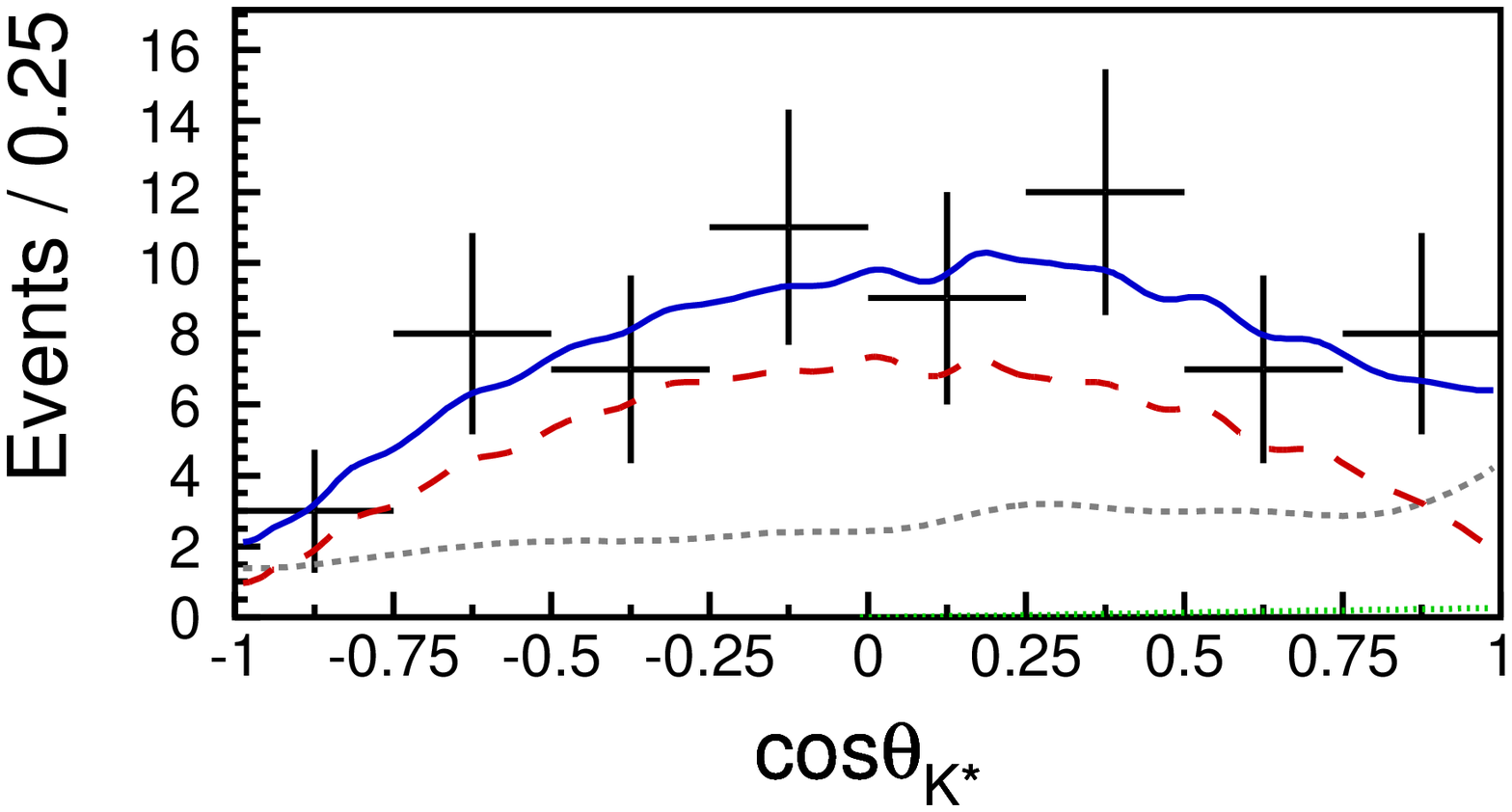} 
\hskip -0.5cm
\includegraphics[width=4.5cm,height=3.5cm]{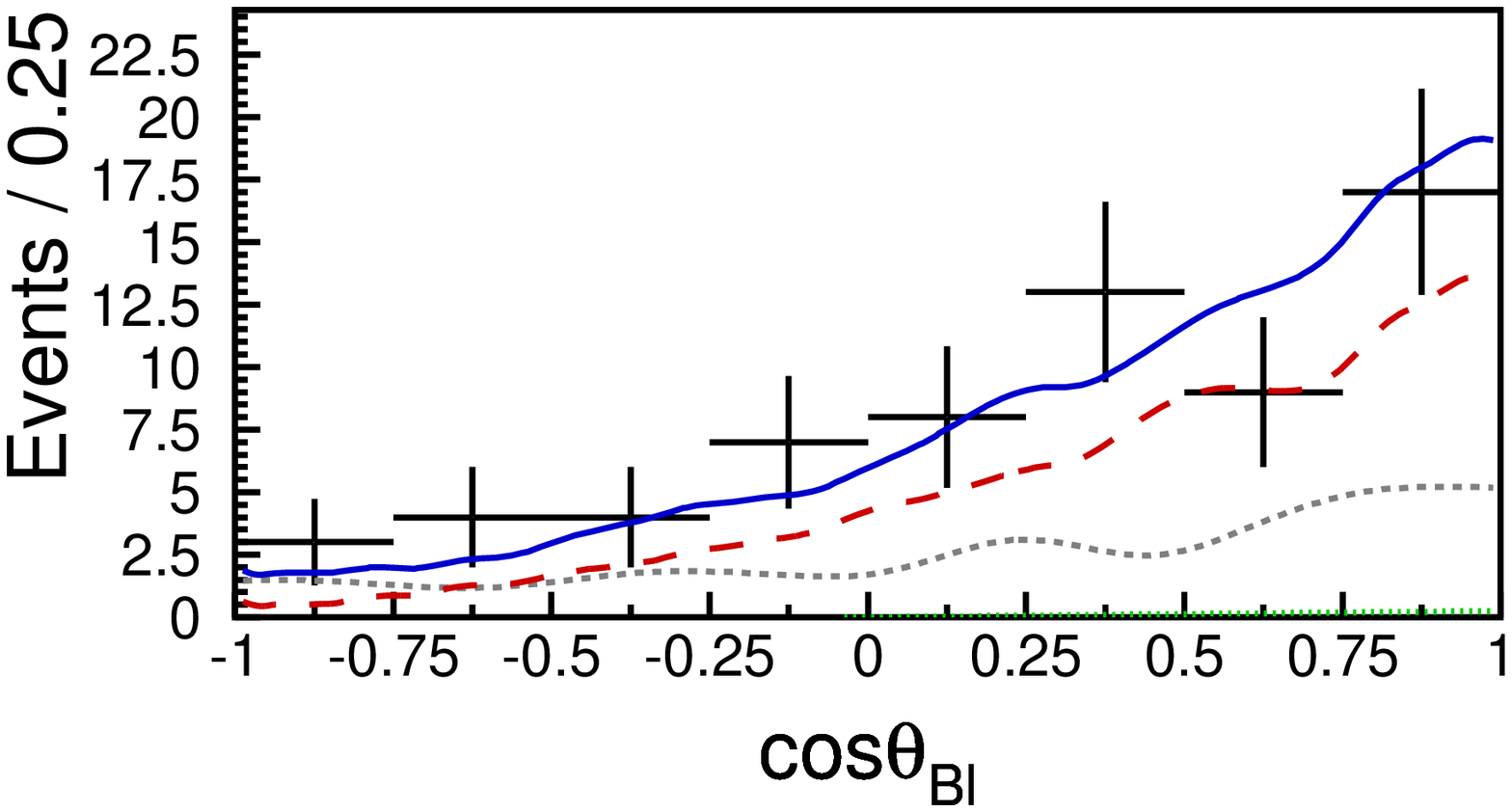} \\
\vskip -3.6cm \hskip -6.4cm {\bf (u)} \hskip 2.7cm {\bf (v)} \\
\vskip -0.1cm \hskip 10.2cm {\bf (w)} \hskip 3.4cm {\bf (x)} \\
\vskip 2.2cm
\end{center}
\vskip 0.4cm
\caption{
Fits to $M_{K\pi}$, $M_{\rm bc}$, $\cos \theta_{K^*}$, and $\cos \theta_{B\ell}$ 
for the $K^{*}\ell^+\ell^-$ decays in 6 $q^2$ (GeV$^2$/$c^2$) bins: 
(a)$\sim$(d)~0.00--2.00, (e)$\sim$(h)~2.00--4.30, (i)$\sim$(l)~4.30--8.68, 
(m)$\sim$(p)~10.09--12.86, (q)$\sim$(t)~14.18--16.00, and (u)$\sim$(x)~$>16.00$.
The solid, long-dashed, short-dashed, and dotted curves represent 
the combined fit result, fitted signal, combinatorial background, 
and $J/\psi(\psi^\prime) X$ background, respectively.
}
\label{fig:kstllfit}
\end{figure}

\begin{figure}[htb]
\begin{center}
\vskip -0.5cm 
\includegraphics[width=3.8cm]{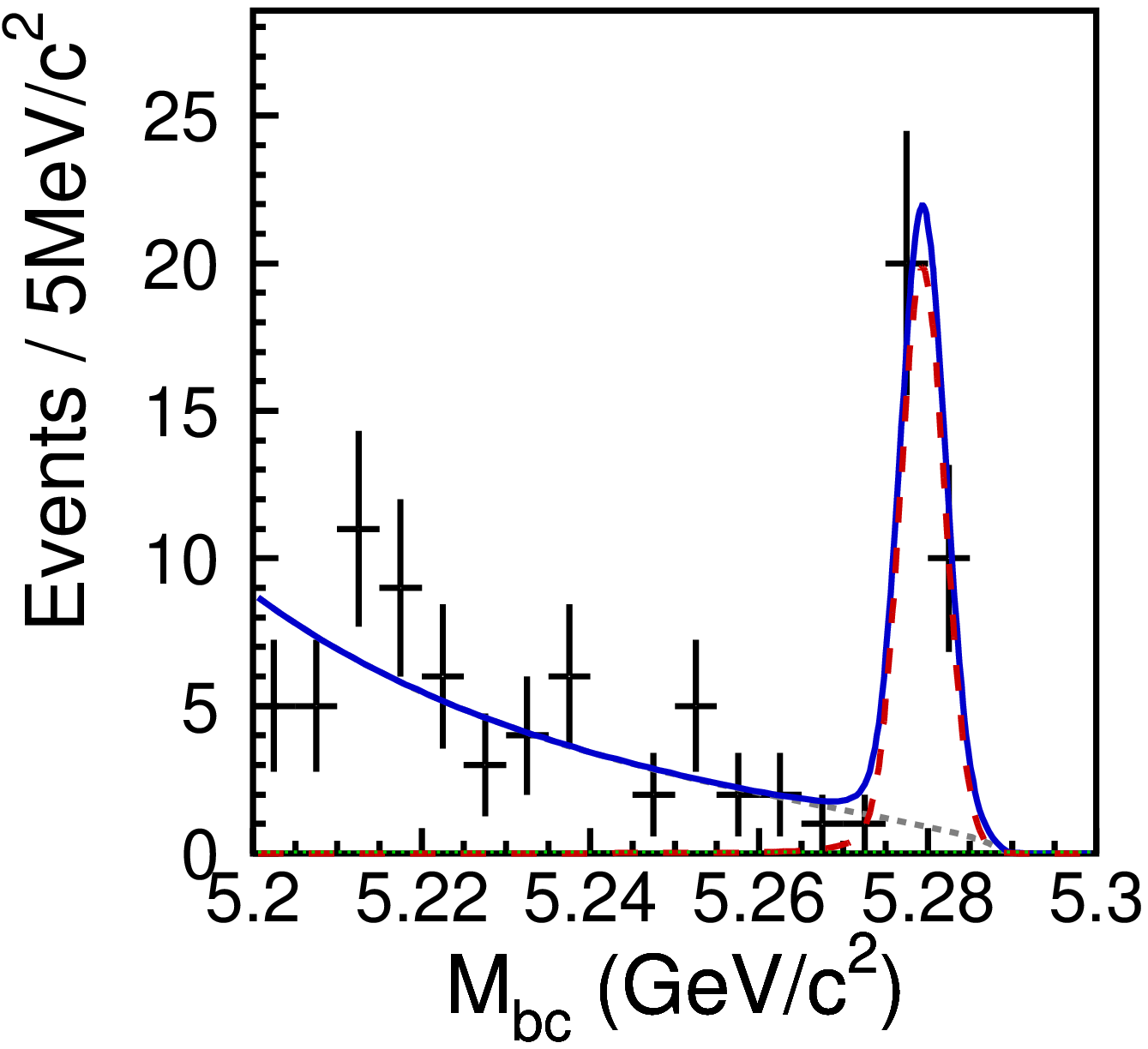} 
\hskip -0.5cm
\includegraphics[width=4.5cm,height=4.0cm]{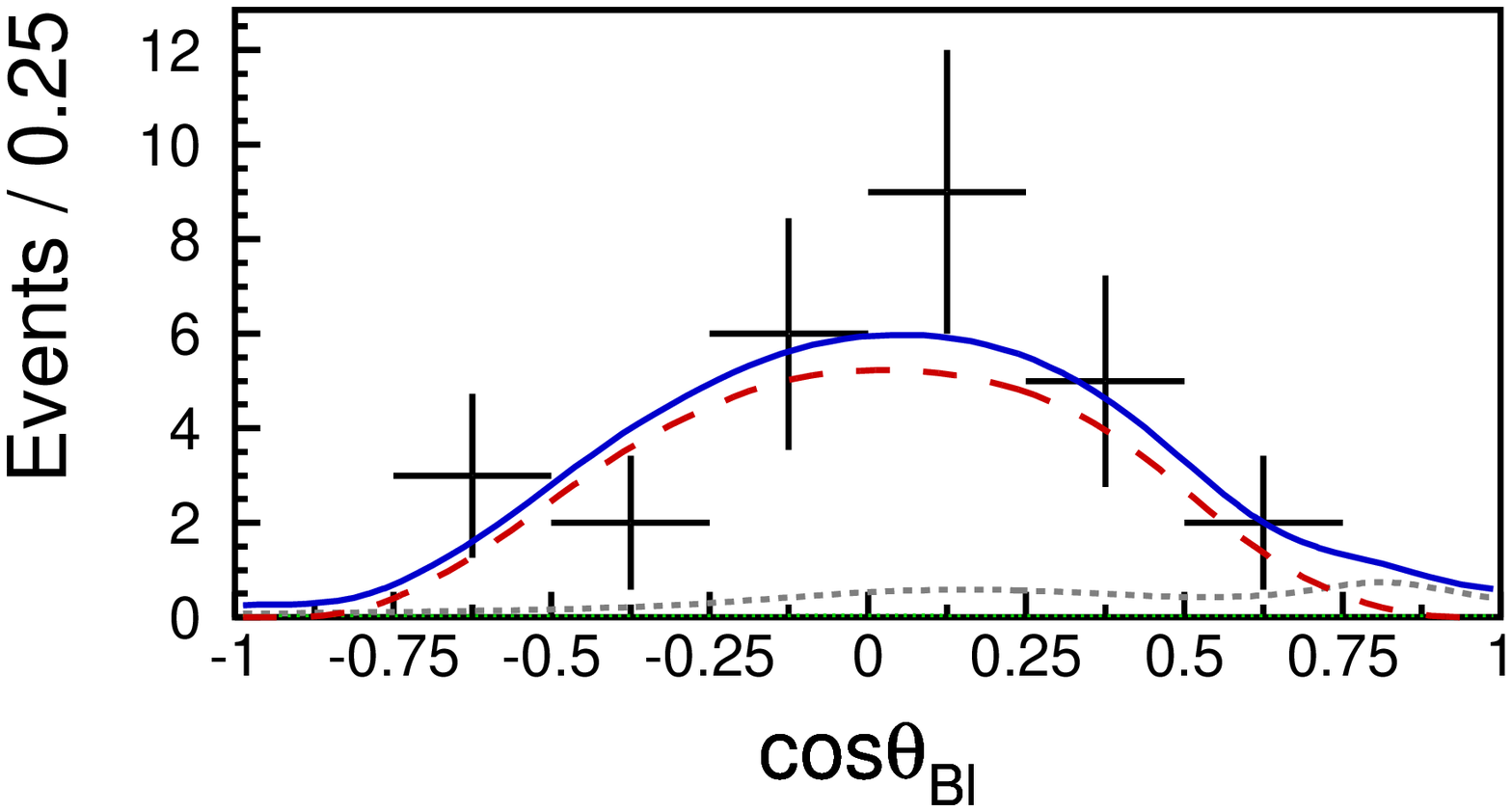} 
\hskip 0.5cm
\includegraphics[width=3.8cm]{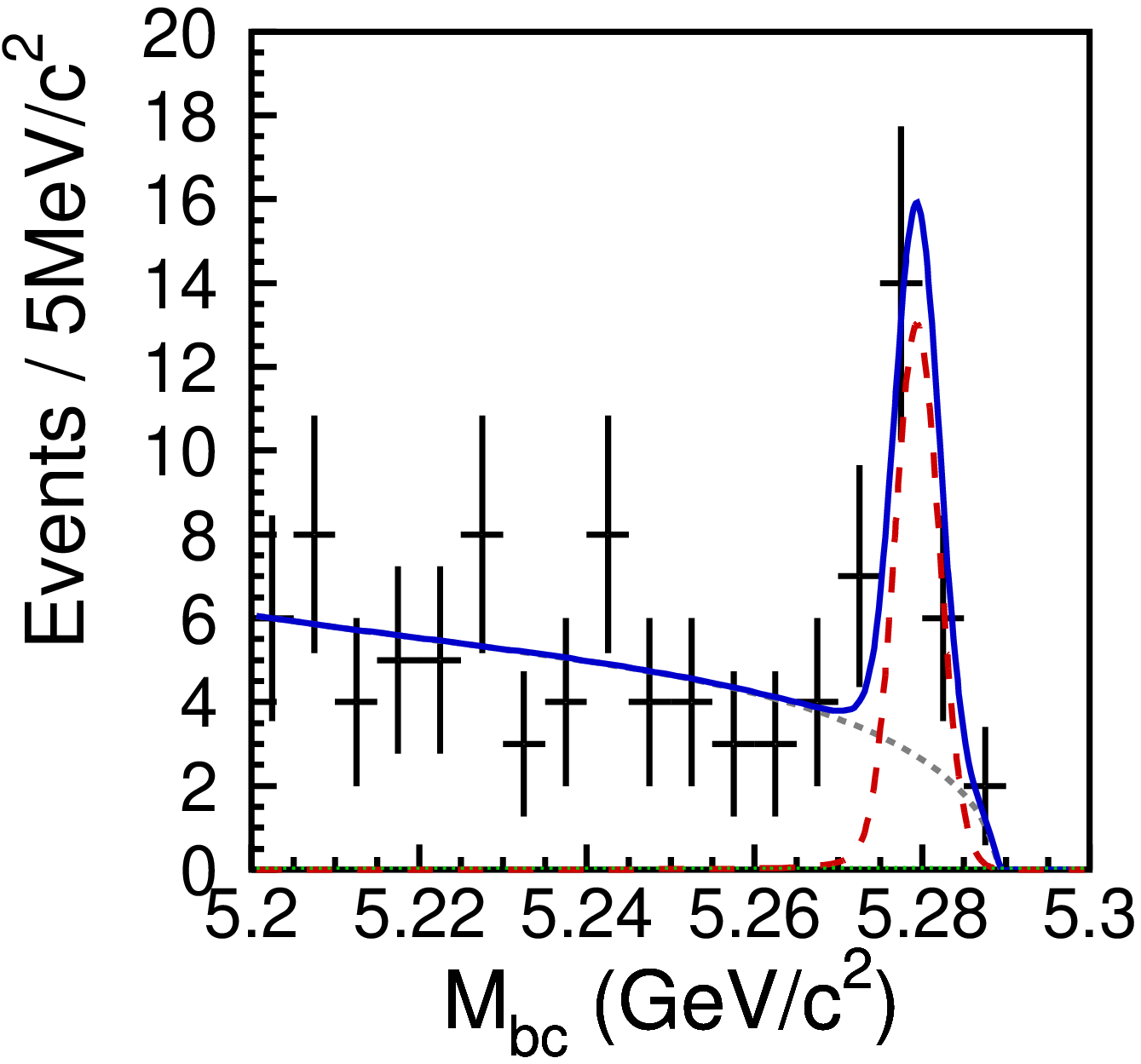} 
\hskip -0.5cm
\includegraphics[width=4.5cm,height=4.0cm]{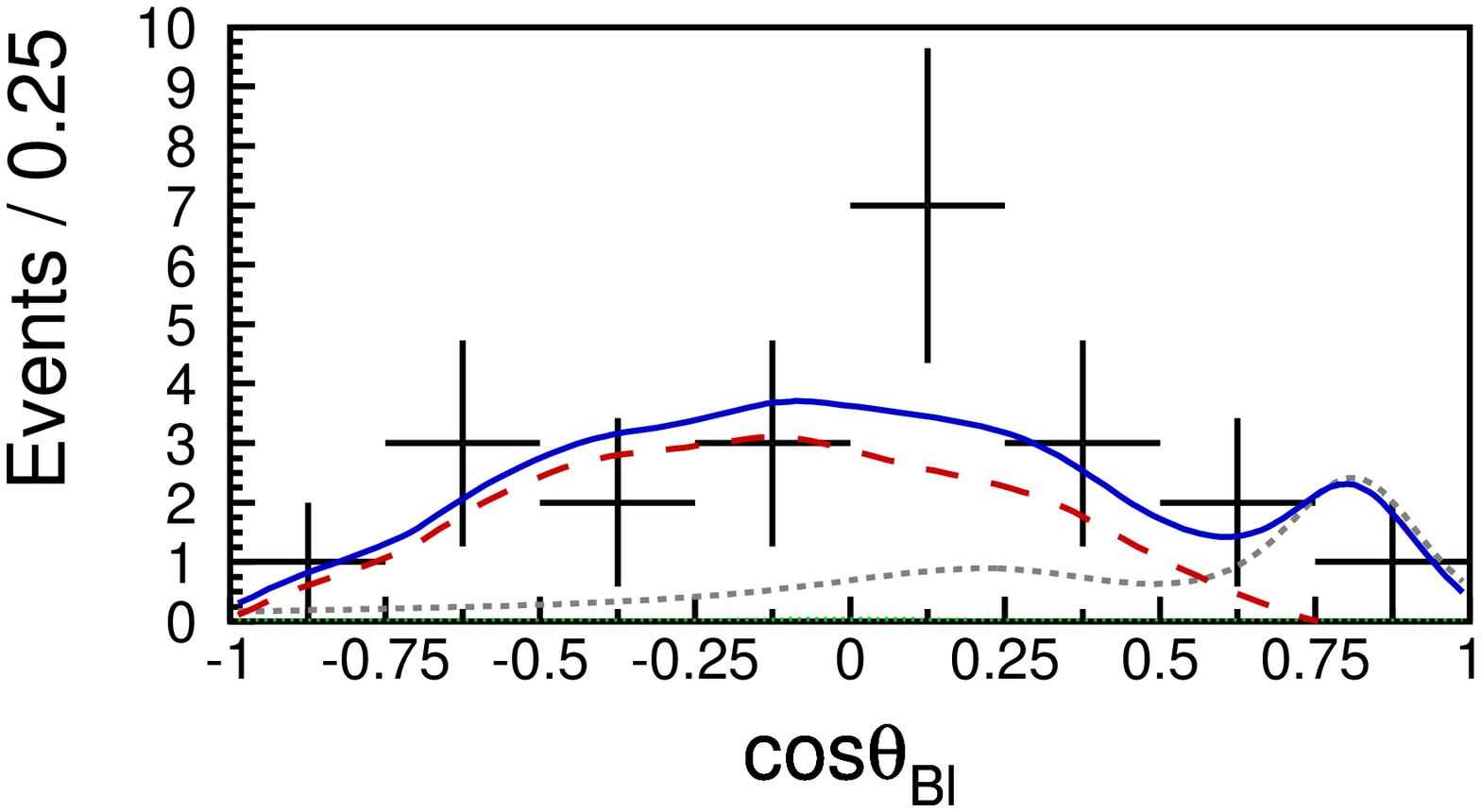} \\
\vskip -3.5cm 
\hskip 2.2cm {\bf (a)} \hskip 3.4cm {\bf (b)} 
\hskip 3.7cm {\bf (c)} \hskip 3.4cm {\bf (d)}\\
\vskip 2.2cm
\includegraphics[width=3.8cm]{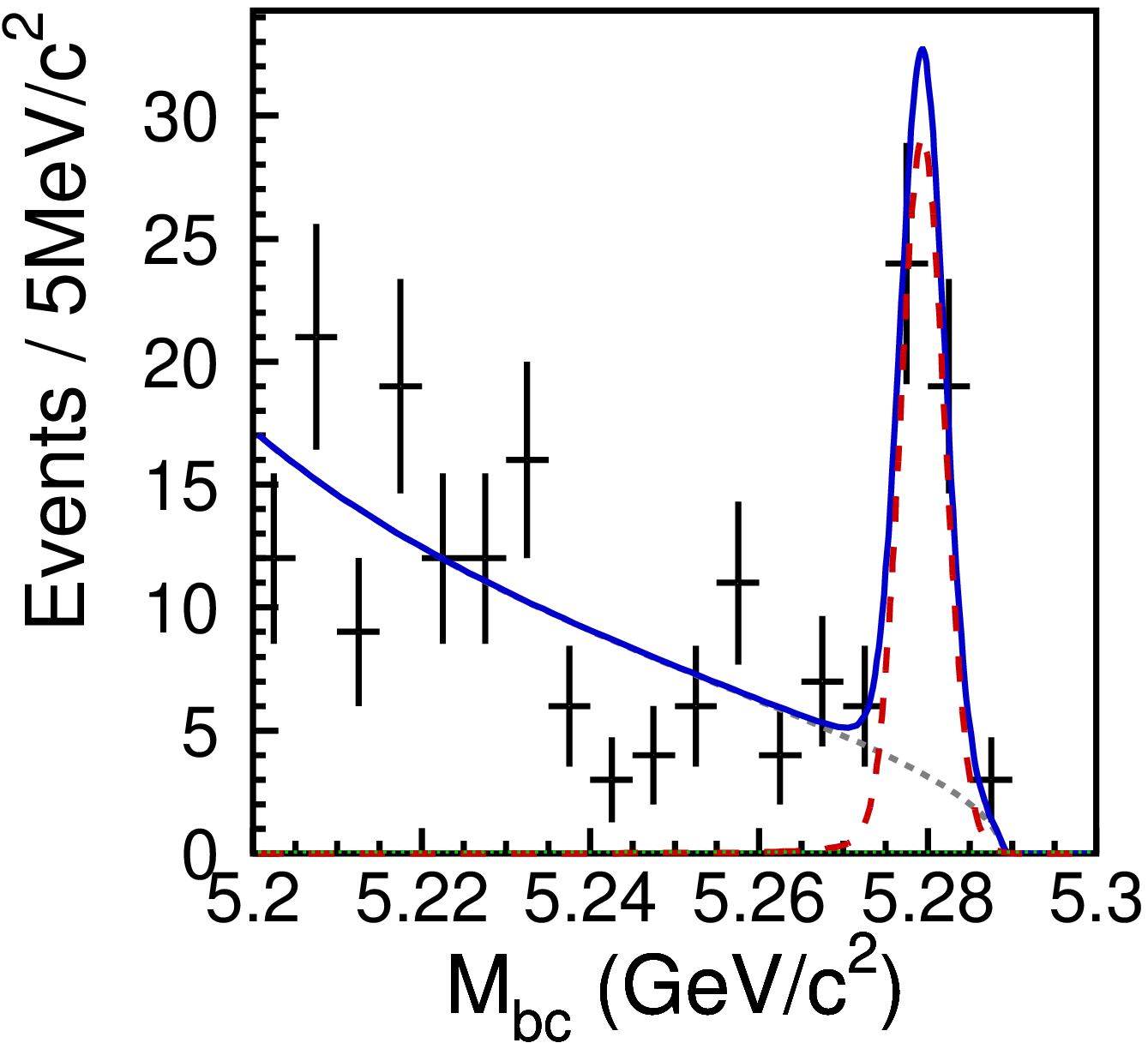} 
\hskip -0.5cm
\includegraphics[width=4.5cm,height=4.0cm]{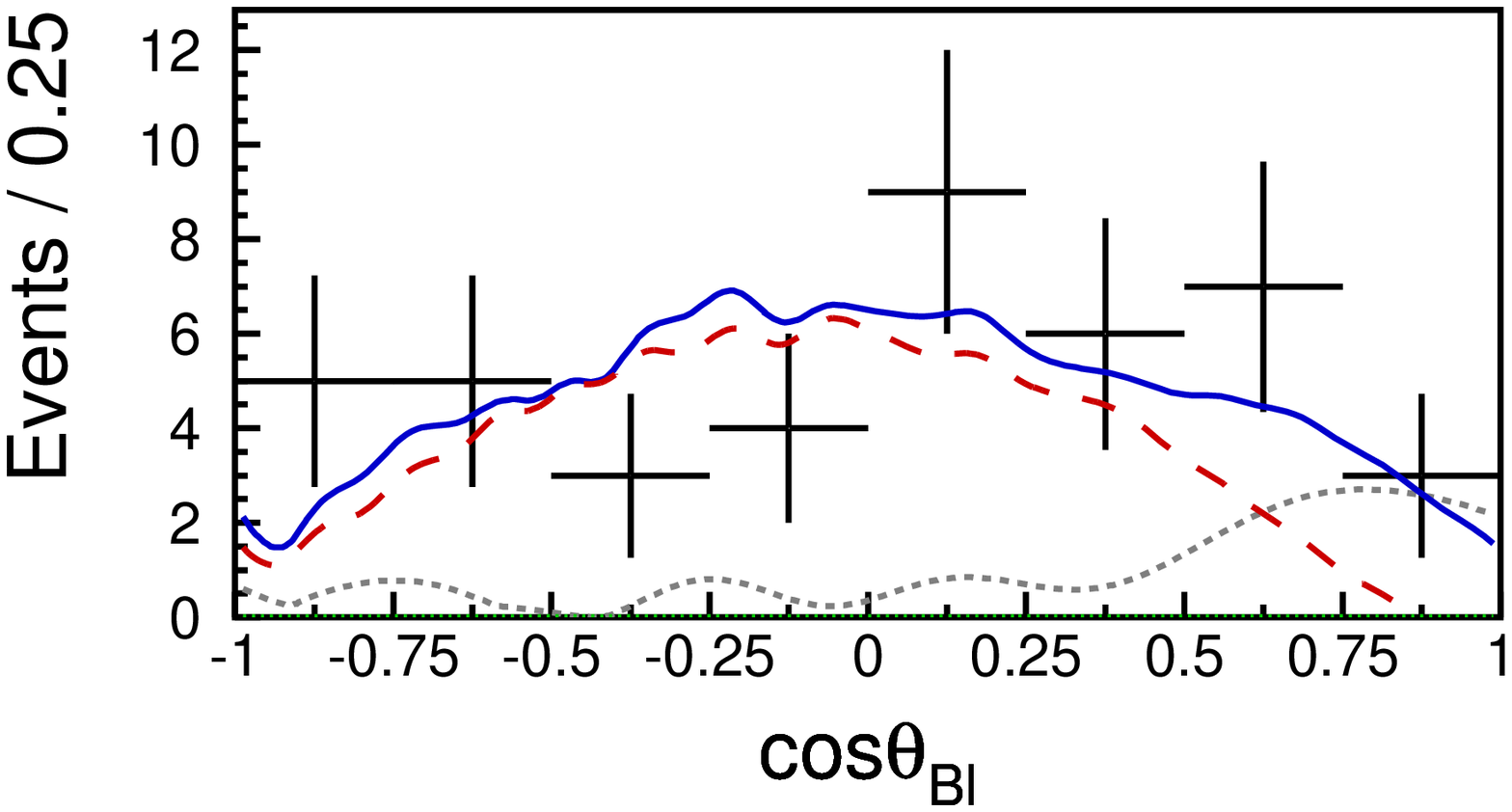} 
\hskip 0.5cm
\includegraphics[width=3.8cm]{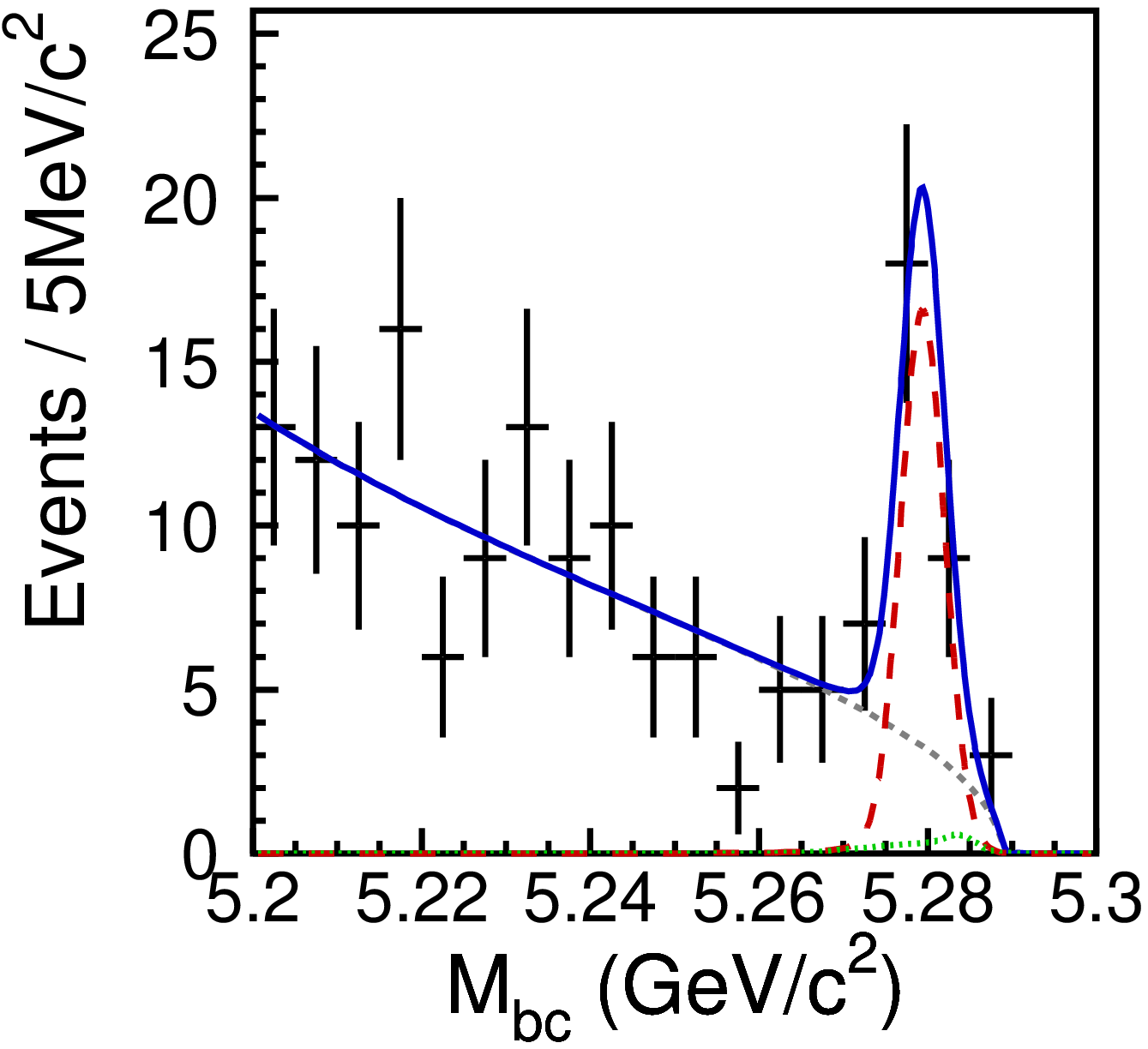} 
\hskip -0.5cm
\includegraphics[width=4.5cm,height=4.0cm]{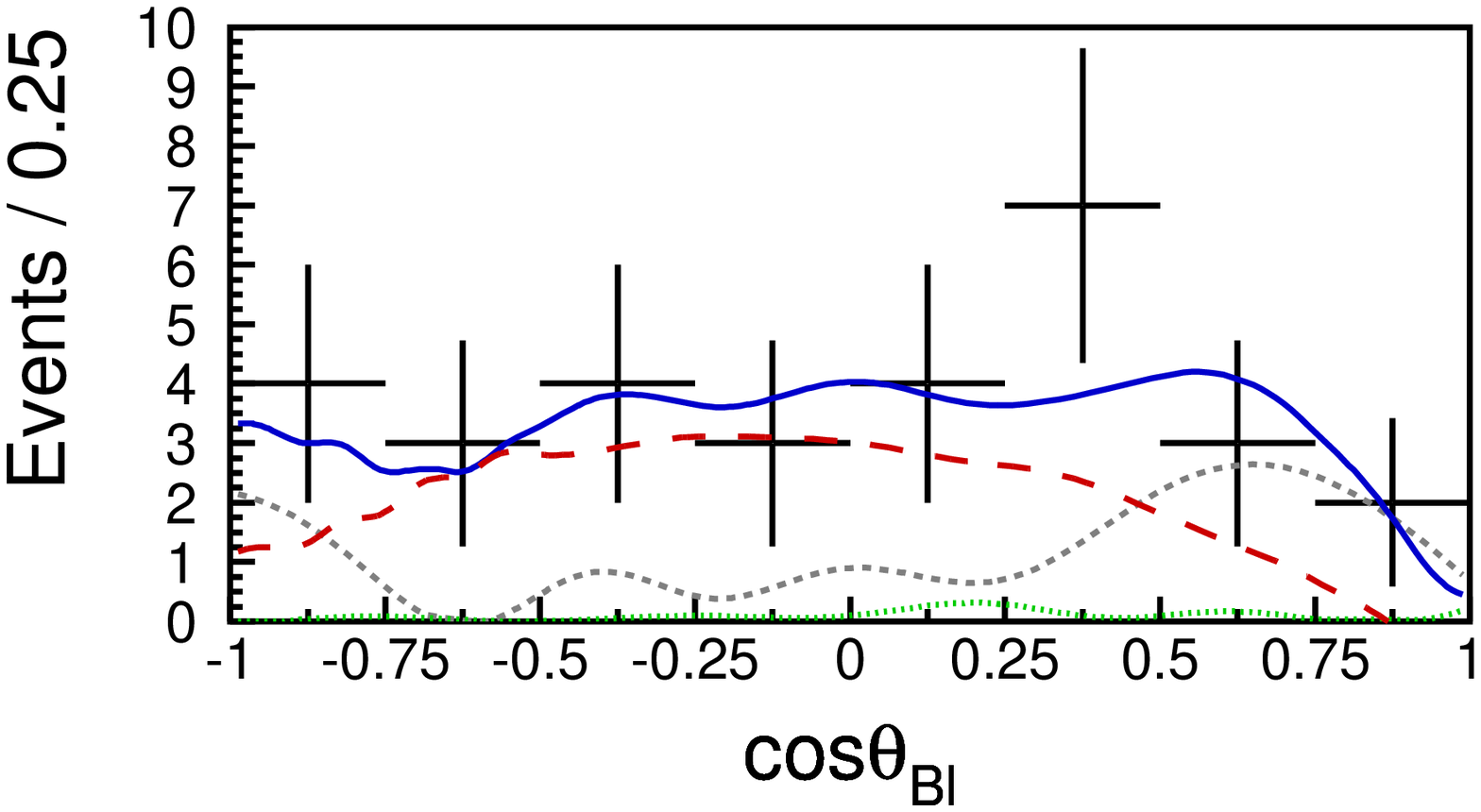} \\
\vskip -3.5cm 
\hskip 2.2cm {\bf (e)} \hskip 3.4cm {\bf (f)} 
\hskip 3.7cm {\bf (g)} \hskip 3.4cm {\bf (h)}\\
\vskip 2.2cm
\includegraphics[width=3.8cm]{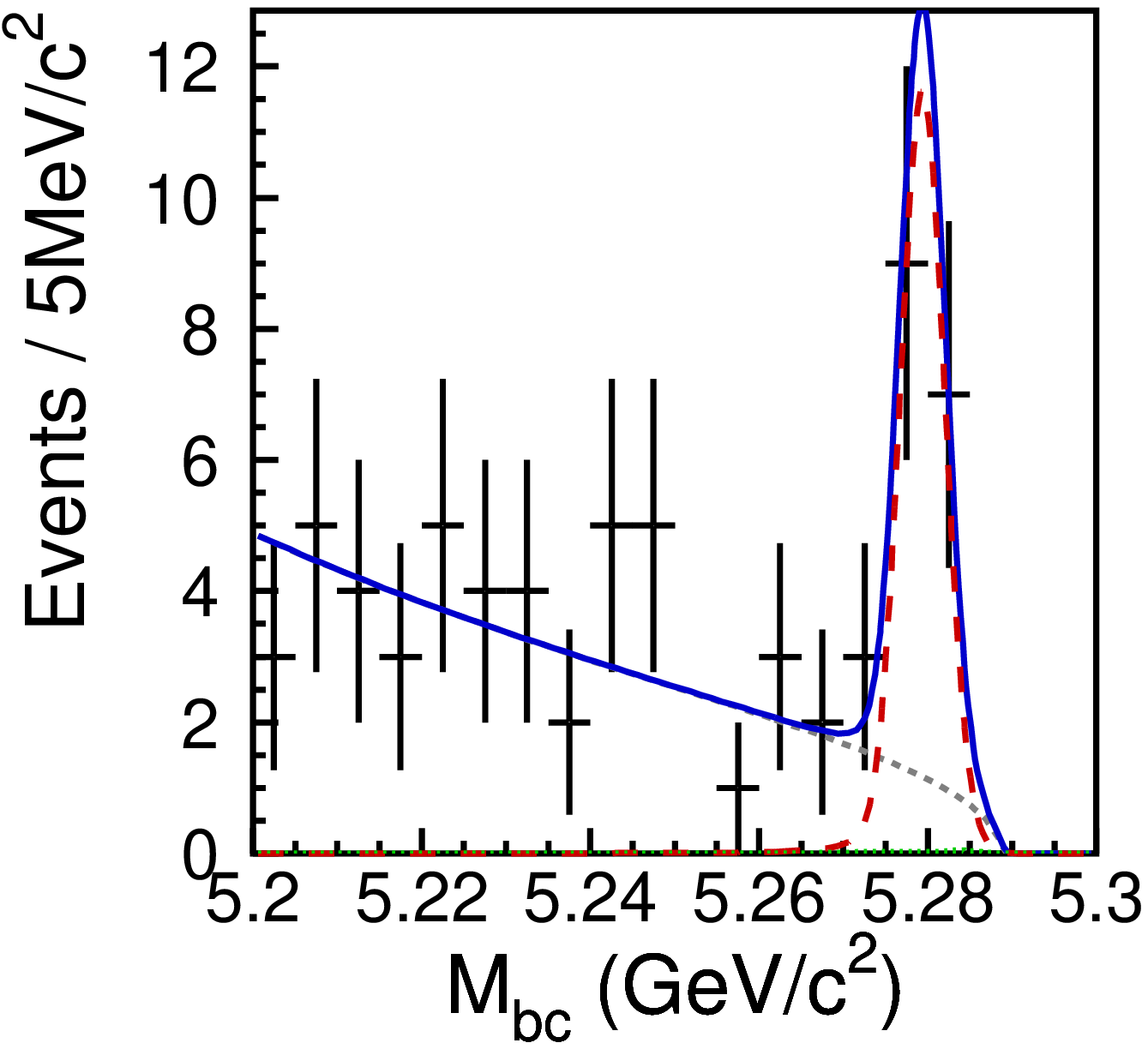} 
\hskip -0.5cm
\includegraphics[width=4.5cm,height=4.0cm]{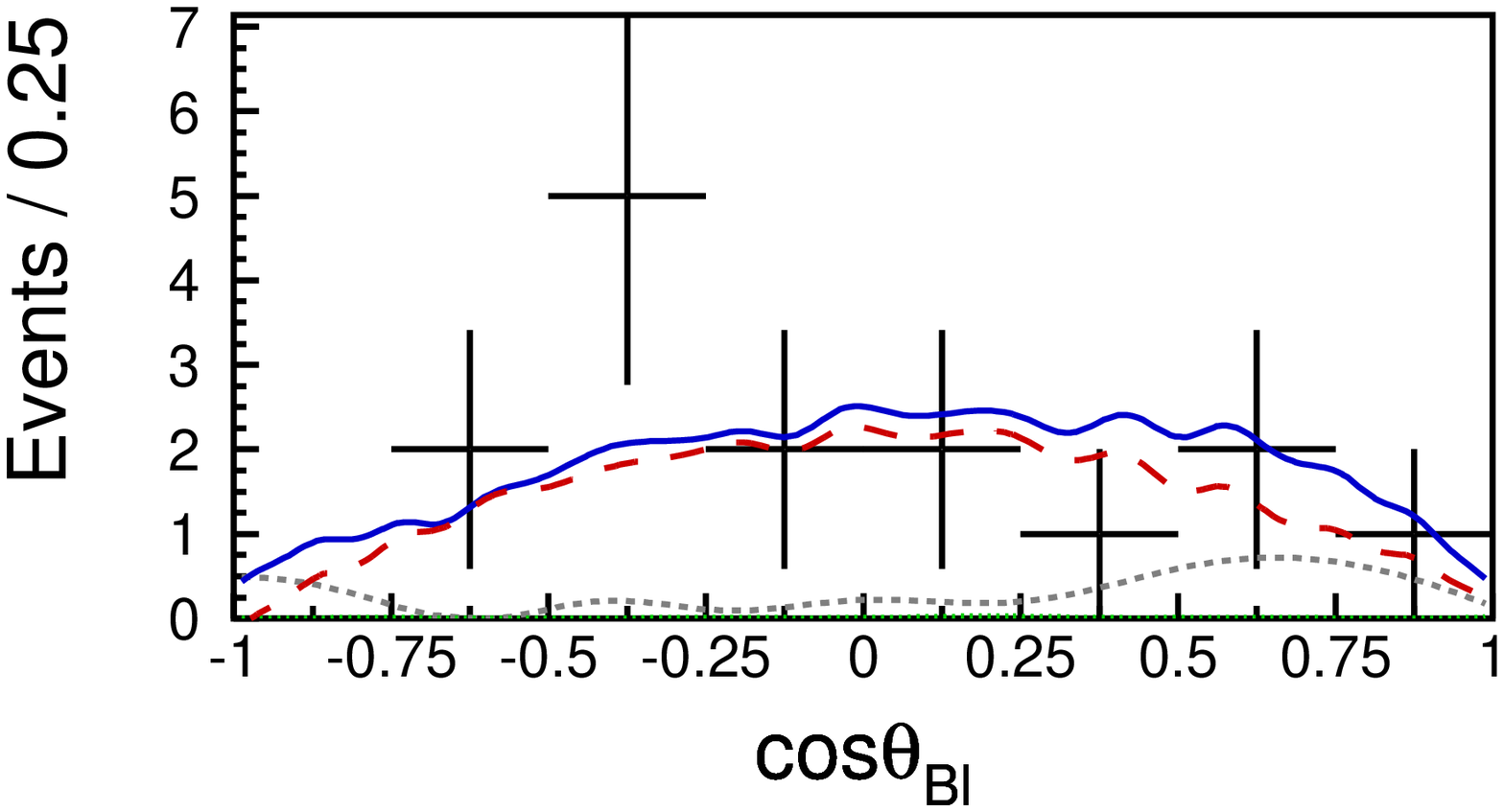} 
\hskip 0.5cm
\includegraphics[width=3.8cm]{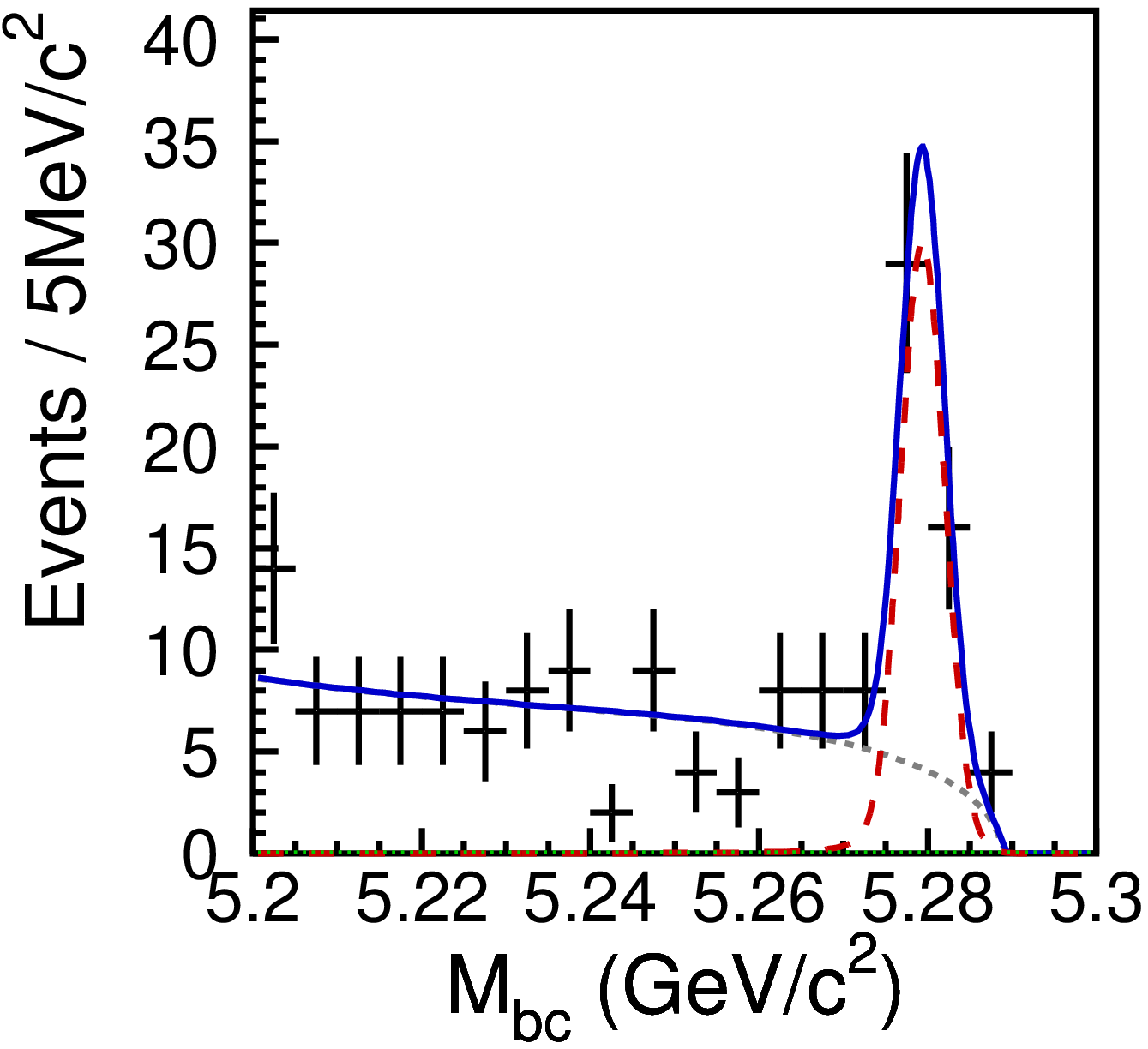} 
\hskip -0.5cm
\includegraphics[width=4.5cm,height=4.0cm]{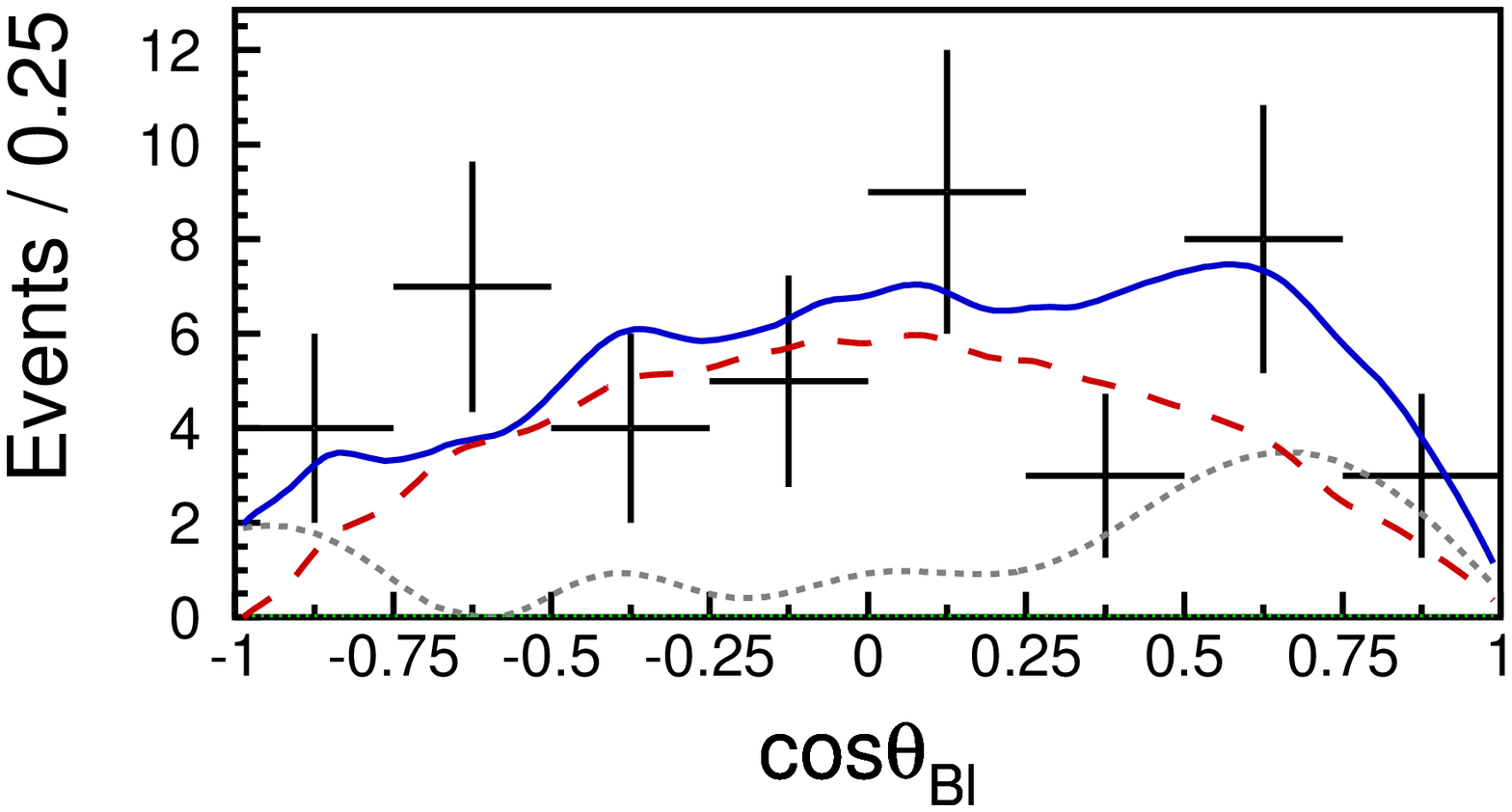} \\
\vskip -3.5cm 
\hskip 2.2cm {\bf (i)} \hskip 3.4cm {\bf (j)} 
\hskip 3.7cm {\bf (k)} \hskip 3.4cm {\bf (l)}\\
\vskip 2.2cm
\end{center}
\vskip 0.4cm
\caption{
Fits to $M_{\rm bc}$ and $\cos \theta_{B\ell}$ 
for the $K\ell^+\ell^-$ decays in 6 $q^2$ (GeV$^2$/$c^2$) bins: 
(a)$\sim$(b)~0.00--2.00, (c)$\sim$(d)~2.00--4.30, (e)$\sim$(f)~4.30--8.68, 
(g)$\sim$(h)~10.09--12.86, (i)$\sim$(j)~14.18--16.00, and (k)$\sim$(l)~$>16.00$.
The solid, long-dashed, and short-dashed represent 
the combined fit result, fitted signal, and combinatorial background, respectively. 
The$J/\psi(\psi^\prime) X$ background is so small and not represented. 
}
\label{fig:kllfit}
\end{figure}

\begin{table}[htb]
\caption{
Total branching fractions and $CP$ asymmetries for subsamples of  
$B\to K^* \ell^+ \ell^-$ and $B\to K\ell^+\ell^-$ decays. 
}
\label{tb:tbf}
\begin{center}
\begin{tabular}{c|cc}
\hline
\hline
Mode & ${\mathcal B}$ (10$^{-7}$)  & $A_{CP}$  \\
\hline
$K^{*+}\mu^+\mu^-$ & $11.1^{+3.2}_{-2.7}$$\pm$1.0~ & $-0.12^{+0.24}_{-0.24}$$\pm$0.02 \\
$K^{*0}\mu^+\mu^-$ & $10.6^{+1.9}_{-1.4}$$\pm$0.7~ & $0.00^{+0.15}_{-0.15}$$\pm$0.03 \\
$K^{*}\mu^+\mu^-$  & $11.0^{+1.6}_{-1.4}$$\pm$0.8~ & $-0.03^{+0.13}_{-0.13}$$\pm$0.02 \\
\hline
$K^{*+}e^+e^- $ & $17.3^{+5.0}_{-4.2}$$\pm$2.0~ & $-0.14^{+0.23}_{-0.22}$$\pm$0.02 \\
$K^{*0}e^+e^- $ & $11.8^{+2.7}_{-2.2}$$\pm$0.9~ & $-0.21^{+0.19}_{-0.19}$$\pm$0.02 \\
$K^{*}e^+e^- $  & $13.9^{+2.3}_{-2.0}$$\pm$1.2~ & $-0.18^{+0.15}_{-0.15}$$\pm$0.01 \\
\hline
$K^{*+}\ell^+\ell^-$ & $12.4^{+2.3}_{-2.1}$$\pm$1.3~ & $-0.13^{+0.17}_{-0.16}$$\pm$0.01 \\
$K^{*0}\ell^+\ell^-$ &  $9.7^{+1.3}_{-1.1}$$\pm$0.7~ & $-0.08^{+0.12}_{-0.12}$$\pm$0.02 \\
$K^{*}\ell^+\ell^-$  & $10.7^{+1.1}_{-1.0}$$\pm$0.9~ & $-0.10^{+0.10}_{-0.10}$$\pm$0.01 \\
\hline
\hline
$K^{+}\mu^+\mu^-$ & $5.3^{+0.8}_{-0.7}$$\pm$0.3~ & $-0.05^{+0.13}_{-0.13}$$\pm$0.03 \\
$K^{0}\mu^+\mu^-$ & $4.4^{+1.3}_{-1.1}$$\pm$0.3~ & $-$ \\
$K\mu^+\mu^-$     & $5.0^{+0.6}_{-0.6}$$\pm$0.3~ & $-$ \\
\hline
$K^{+}e^+e^- $ & $5.7^{+0.9}_{-0.8}$$\pm$0.3~ & $0.14^{+0.14}_{-0.14}$$\pm$0.03 \\
$K^{0}e^+e^- $ & $2.0^{+1.4}_{-1.0}$$\pm$0.1~ &  $-$ \\
$Ke^+e^- $     & $4.8^{+0.8}_{-0.7}$$\pm$0.3~ &  $-$ \\
\hline
$K^{+}\ell^+\ell^-$ & $5.3^{+0.6}_{-0.5}$$\pm$0.3~ & $0.04^{+0.10}_{-0.10}$$\pm$0.02 \\
$K^{0}\ell^+\ell^-$ & $3.4^{+0.9}_{-0.8}$$\pm$0.2~ & $-$ \\
$K\ell^+\ell^-$     & $4.8^{+0.5}_{-0.4}$$\pm$0.3~ & $-$ \\
\hline
\hline
\end{tabular}
\end{center}
\end{table} 


\begin{thebibliography}{99}
\bibitem{chrg_conj}{
Charge-conjugate decays are implied and 
equal production of $B^0\overline{B}{}^0$ and $B^+B^-$ pairs is assumed 
throughout the paper.
}
\bibitem{error}{
The first and second errors are statistical and systematic, 
respectively, in all measurements quoted throughout the paper.
}
\end{thebibliography}
\end{document}